\documentclass[11pt]{article}

\usepackage[utf8]{inputenc}
\usepackage[T1]{fontenc}
\usepackage{lmodern}
\usepackage{microtype}
\usepackage{amsmath,amssymb}
\usepackage{geometry}
\usepackage{graphicx}
\graphicspath{{./}{Images/}}
\usepackage{booktabs,longtable,tabularx,array,multirow}
\usepackage{caption}
\usepackage{enumitem}
\usepackage{natbib}
\usepackage{hyperref}
\usepackage{xcolor}
\usepackage{float}

\hypersetup{
    colorlinks=true,
    linkcolor=blue!50!black,
    citecolor=blue!50!black,
    urlcolor=blue!50!black,
    pdftitle={Benchmarking Pretrained Time-Series Foundation Models for Financial Return Forecasting}
}

\begin{document}

\begin{center}
{\LARGE\bfseries Pretrained Time-Series Foundation Models for Financial Return Forecasting\par}
\vspace{1em}
\mbox{Miquel Noguer i Alonso$^{1}$\hspace{1.5cm}Rodolfo Pereira Franklin$^{1}$}\par
\vspace{0.4em}
{\small\centering $^{1}$Artificial Intelligence Finance Institute\par}
\vspace{0.4em}
{\today\par}
\end{center}

\vspace{1em}
\begin{abstract}
Financial return forecasting is a difficult test case for time-series foundation models because daily equity returns combine low signal-to-noise ratios, structural breaks, heavy tails, and weak persistence. This paper benchmarks pretrained time-series foundation models (TSFMs) against train-from-scratch neural baselines in a deliberately conservative financial setting. We evaluate TimeGPT/TimeGPT-LH, TimesFM-2.5, Moirai-2.0, Chronos, and Chronos-2 against NBEATS, NHITS, PatchTST, iTransformer, and KAN on five liquid U.S. equities (AAPL, AMZN, GOOG, JPM, and META), using both linear and log returns. All neural models are compared under an equalized context budget of $L=512$ observations, a rolling-origin protocol, and error metrics benchmarked against naive random-walk alternatives. The paper also provides a compact theoretical framing of pretraining as an inductive prior, linking PAC-Bayes transfer intuition, information-theoretic predictability limits, attention geometry, and score-distribution comparisons to the empirical design. This framing clarifies why strong rankings need not imply economically meaningful predictability in noisy markets. The main finding is pragmatic rather than sensational. Pretrained TSFMs dominate the ranking distribution, accounting for 8 of 10 task-level wins, with Moirai-2.0 and TimesFM-2.5 achieving the strongest average ranks. TimesFM-2.5 leads the AAPL and JPM tasks, Moirai-2.0 leads the GOOG tasks and one AMZN task, and Chronos wins the remaining AMZN task. However, the iTransformer baseline wins both META tasks, showing that local supervised learning can still outperform generic pretraining for specific assets and regimes. More importantly, gains over the random-walk benchmark are small and sparse. A one-sided Diebold--Mariano test rejects equal or inferior predictive accuracy only for Chronos on AMZN and Moirai-2.0 on GOOG. The evidence therefore supports TSFMs as useful practical priors that reduce model-development costs in low-data financial forecasting, but not as universal engines of statistically reliable alpha generation or trading performance in realistic empirical deployment conditions.
\end{abstract}

\noindent\textbf{Keywords:} Time-series foundation models; financial return forecasting; pretrained forecasting models; equity returns; rolling-origin evaluation; Diebold--Mariano test; TimeGPT; TimesFM; Chronos; Moirai; neural forecasting.

\newpage
\tableofcontents
\newpage

\section{Introduction}

Time series analysis has long been a fundamental pillar of stochastic modelling,
with its formal foundations in autoregressive models dating back to Yule and Walker
in the 1920s and 1930s, and taking its modern form with the systematic ARMA
framework of \citet{boxjenkins1970timeseriesanalysisforecastingandcontrol}.
Forecasting financial returns within this tradition is notoriously difficult:
signal-to-noise ratios are low, structural breaks are common, and even modest errors
can erase any practical edge once trading frictions are considered. Recent progress
in pretrained sequence models has nevertheless renewed interest in whether
large-scale temporal pretraining can provide a better inductive bias than training
a separate model for each asset.

This literature now contains two related but distinct strands. One strand adapts
text large language models (LLMs) to time series through reprogramming or prompting,
as in Time-LLM \citep{jin2024timellmtimeseriesforecasting}, LLM4TS
\citep{chang2024llm4tsaligningpretrainedllms}, and PromptCast
\citep{xue2023promptcastnewpromptbasedlearning}. As emphasized by
\citet{brown2020languagemodelsfewshotlearners}, LLMs such as GPT-4 \citep{openai2023}
and Gemini \citep{geminiteam2024geminifamilyhighlycapable} exhibit remarkable
transfer-learning capabilities in few-shot and zero-shot settings, motivating the broader hypothesis that pretrained sequence intelligence may transfer to temporal data. Building on our previous findings regarding the efficacy of general-purpose Large Language Models adapted for financial time series through reprogramming techniques \citep{AlonsoFranklin2025}, this benchmark aims to extend that inquiry into the emerging class of models natively designed for temporal data. A second strand develops dedicated time-series foundation models (TSFMs),
including TimeGPT \citep{garza2024timegpt1}, TimesFM
\citep{das2024decoderonlyfoundationmodeltimeseries},
Chronos \citep{ansari2024chronoslearninglanguagetime}, Chronos-2
\citep{ansari2025chronos2}, and Moirai-2.0 \citep{liu2026moirai20timeseries}.
The present benchmark belongs much more clearly to the second strand: most evaluated
systems are pretrained time-series models rather than reprogrammed text LLMs. For
that reason, this paper uses the term \emph{pretrained TSFM} throughout.

The question studied here is practical. If a practitioner has only one return series
per asset, wants a 20-business-day forecast, and does not want to run heavy
per-ticker model development, is an out-of-the-box TSFM preferable to a
train-from-scratch alternative? To answer that question, we compare six pretrained
TSFMs against five strong baselines: NBEATS
\citep{oreshkin2020nbeatsneuralbasisexpansion},
NHITS \citep{challu2022nhitsneuralhierarchicalinterpolation},
PatchTST \citep{nie2023timeseriesworth64},
iTransformer \citep{liu2024itransformerinvertedtransformerseffective},
and KAN \citep{xu2024kolmogorovarnoldnetworkstimeseries}.

The contribution of the paper is threefold. First, it provides a compact benchmark
over two return representations, five liquid equities, and a rolling-origin
protocol, with detailed architectural exposition of each evaluated model. Second,
it reframes the evidence around pretrained time-series models rather than generic
``LLM'' claims. Third, it separates statistical distinguishability from directional
superiority when interpreting Diebold--Mariano (DM) results, preventing sign
confusion relative to the random-walk benchmark.

\section{Related Work}

A related but distinct line of work adapts frozen text LLMs to time-series data
through prompting or reprogramming, as in Time-LLM
\citep{jin2024timellmtimeseriesforecasting}, LLM4TS
\citep{chang2024llm4tsaligningpretrainedllms}, and PromptCast
\citep{xue2023promptcastnewpromptbasedlearning}. This literature motivates the
broader hypothesis that sequence pretraining can transfer across modalities, but
it is not the empirical focus of this paper: we do not run a direct Time-LLM or
LLM-reprogramming experiment.

The pretrained-TSFM literature is closer to the systems evaluated here. TimeGPT
\citep{garza2024timegpt1} emphasizes transfer learning for zero-shot forecasting.
TimesFM is a decoder-only foundation model for time-series forecasting
\citep{das2024decoderonlyfoundationmodeltimeseries}. Chronos casts forecasting as a
language modeling problem over quantized values
\citep{ansari2024chronoslearninglanguagetime}, while Chronos-2 extends that
philosophy to a broader universal-forecasting setting with multivariate and
covariate support \citep{ansari2025chronos2}. Moirai-2.0 is a decoder-only
foundation model built around quantile forecasting and efficiency improvements
\citep{liu2026moirai20timeseries}.

A complementary line of contemporaneous work asks the related but distinct
question of how far generic TSFM pretraining transfers to finance-specific
data. \citet{rahimikia2025revisiting} evaluate zero-shot inference,
fine-tuning, and pre-training from scratch for several TSFM families on a broad
global panel of excess-return observations. Their main finding is that
off-the-shelf TSFMs underperform standard ensemble and neural benchmarks in
zero-shot and fine-tuning regimes, while finance-native pretraining closes much
of the gap and can produce stronger portfolio performance. Our benchmark differs
in scope---single-asset 20-business-day forecasts on five U.S. equities, rather
than next-day cross-sectional excess returns at the global level---but the two
studies are mutually informative. Theirs argues that domain-specific pretraining
matters when the goal is cross-sectional return prediction and portfolio
formation; ours measures how the same off-the-shelf TSFMs fare against
train-from-scratch deep baselines for per-asset forecast error. The qualitative
messages are consistent: gains from generic TSFMs can be useful in a low-data
per-asset setting, but they remain limited and should not be confused with
finance-specific economic outperformance.

Against these pretrained models, the paper uses strong supervised baselines. NBEATS
and NHITS remain influential MLP-style decomposition models
\citep{oreshkin2020nbeatsneuralbasisexpansion,challu2022nhitsneuralhierarchicalinterpolation}.
PatchTST and iTransformer are modern transformer baselines that model temporal
dependencies in different ways
\citep{nie2023timeseriesworth64,liu2024itransformerinvertedtransformerseffective}.
KAN brings spline-based Kolmogorov--Arnold parameterization to time-series
prediction \citep{xu2024kolmogorovarnoldnetworkstimeseries} and proves particularly
relevant here because it is the only train-from-scratch model that consistently
breaks into the top tier.

\section{Methodology}
\label{sec:methodology}

This section collects the architectural background of the evaluated systems. It
begins with a shared problem formulation, then describes each model individually.
The exposition follows the canonical references for each architecture and is
provided to make the benchmark self-contained.

\subsection{Problem Formulation}
\label{sec:problem-formulation}

Fix a probability space $(\Omega, \mathcal{F}, \mathbb{P})$ supporting a real-valued
return process $(r_t)_{t \in \mathbb{Z}}$ adapted to its natural filtration
$\mathcal{F}_t = \sigma(r_s : s \le t)$. For an asset return at horizon $h \ge 1$
we seek a measurable forecast
\begin{equation}
\hat r_{t+h} = f_t\!\big(r_{t-L+1}, \ldots, r_t\big),
\end{equation}
based on a context window of length $L$, where $f_t$ ranges over a class
$\mathcal{F}$ of admissible predictors. In our experiments $h$ runs over
$1, \ldots, H$ with $H = 20$. For a strictly proper loss
$\ell : \mathbb{R} \times \mathbb{R} \to \mathbb{R}_{\ge 0}$ the population risk
and the Bayes-optimal predictor are
\begin{equation}
\mathcal{R}(f) = \mathbb{E}\!\left[\ell\!\big(f_t(\mathcal{F}_t),\, r_{t+h}\big)\right],
\qquad
f^\star_{t,h} \in \arg\min_{g \in L^0(\mathcal{F}_t)}
\mathbb{E}\!\left[\ell(g, r_{t+h}) \mid \mathcal{F}_t\right].
\end{equation}
The two losses that organize the analysis below are squared loss and absolute
loss, with qualitatively different optima.

\paragraph{Semimartingale framework.} It is convenient to view $(r_t)$ as the
sequence of increments of a discrete-time process $R_t = \sum_{s \le t} r_s$
adapted to $(\mathcal{F}_t)$. By the Doob decomposition, $R_t = M_t + A_t$
splits uniquely into an $(\mathcal{F}_t)$-martingale $M_t$ with $M_0 = 0$ and
an $(\mathcal{F}_{t-1})$-predictable process $A_t$ of locally bounded variation
\citep{williams1991probability}. The martingale-difference hypothesis
$\mathbb{E}[r_{t+h} \mid \mathcal{F}_t] = 0$ corresponds to $A \equiv 0$, in
which case $r_t = \Delta M_t$ is the increment of a pure martingale. Under
this null the random-walk benchmark introduced below is exactly the
Bayes-optimal $L^2$ predictor; departures from the null---predictable drifts,
regime-switching means, calendar effects---are precisely what any non-trivial
forecaster aims to exploit. Throughout this paper we work without imposing the
martingale-difference null on $(r_t)$: the random walk is treated as a
benchmark to beat, not as a maintained hypothesis.

\paragraph{Squared loss.} For $\ell(\hat r, r) = (\hat r - r)^2$, the
Bayes-optimal predictor is the conditional mean
$f^\star_{t,h}(\mathcal{F}_t) = \mathbb{E}[r_{t+h} \mid \mathcal{F}_t]$, and the
minimum risk equals the expected conditional variance
$\mathbb{E}\,\mathrm{Var}(r_{t+h}\mid\mathcal{F}_t)$. Most foundation-model
training objectives target this loss, either directly (TimesFM, Moirai's
quantile head specializes to a mean-like predictor at $q=0.5$) or via a quantized
likelihood that converges to it under refinement (Chronos, Chronos-2).

\paragraph{Absolute loss.} For $\ell(\hat r, r) = |\hat r - r|$, the
Bayes-optimal predictor is the conditional median
$f^\star_{t,h}(\mathcal{F}_t) = \mathrm{med}(r_{t+h} \mid \mathcal{F}_t)$, and the
minimum risk equals
$\mathbb{E}\big[\,|r_{t+h} - \mathrm{med}(r_{t+h}\mid\mathcal{F}_t)|\,\big]$.
This is the reference for our reported MAE: any model whose conditional forecasts
deviate systematically from the conditional median pays a penalty in MAE that no
amount of mean-targeting can recover. Because most evaluated systems are trained
to minimize a squared- or quantized-likelihood loss rather than absolute loss,
the metric and the training objective are deliberately not aligned, and the
benchmark therefore tests robustness to that misalignment alongside raw
predictive skill.

\paragraph{The random-walk benchmark and two coinciding hypotheses.} The
zero-return random-walk forecast $\hat r^{\mathrm{RW}}_{t+h} \equiv 0$
corresponds, on daily liquid-equity data, to two distinct hypotheses that
happen to coincide. First, under the martingale hypothesis
$\mathbb{E}[r_{t+h}\mid\mathcal{F}_t] = 0$ it is Bayes-optimal under squared
loss. Second, when the unconditional median of $r_{t+h}$ is approximately zero
(a good empirical approximation for the assets and window used here) it is the
unconditional minimizer of $\mathbb{E}|r_{t+h}|$. Beating the random walk in
MAE therefore requires a model to extract genuine \emph{conditional} information
beyond the unconditional median: rescaling or recentering forecasts cannot move
skill from negative to positive without conditional signal.

\paragraph{A predictability bound.} The standard variance decomposition
\begin{equation}
\mathrm{Var}(r_{t+h})
= \mathrm{Var}\!\big(\mathbb{E}[r_{t+h}\mid\mathcal{F}_t]\big)
+ \mathbb{E}\,\mathrm{Var}(r_{t+h}\mid\mathcal{F}_t)
\end{equation}
yields the population upper bound
\begin{equation}
\sup_{f}\; R^2(f)
= \frac{\mathrm{Var}\!\big(\mathbb{E}[r_{t+h}\mid\mathcal{F}_t]\big)}{\mathrm{Var}(r_{t+h})}.
\label{eq:predictability-bound}
\end{equation}
For daily liquid-equity returns this bound is empirically very small: reported
out-of-sample $R^2$ values in the cross-sectional return-prediction literature
are typically below 1\% \citep{rahimikia2025revisiting}. Equivalently, the MAE
ratio between any feasible predictor and the random walk is bounded below by a
quantity close to one. The skill scores of order $10^{-3}$ reported in
Section~\ref{sec:results} should be read against this ceiling rather than
against an unrestricted benchmark.

\paragraph{Empirical risk minimization.} In practice we replace population risk
by an empirical analogue. For a training sample
$\{(\mathbf{x}^{(i)}_{1:L}, \mathbf{y}^{(i)}_{1:H})\}_{i=1}^{n}$ the standard
objective is
\begin{equation}
\widehat{\mathcal{R}}_n(f)
= \frac{1}{n H} \sum_{i=1}^{n} \sum_{h=1}^{H} \ell\!\big(f(\mathbf{x}^{(i)}_{1:L})_h,\, y^{(i)}_h\big),
\label{eq:erm}
\end{equation}
which for the reprogramming framework of \citet{jin2024timellmtimeseriesforecasting}
specializes to the squared-error form
$\widehat{\mathcal{L}} = \tfrac{1}{H}\sum_{h=1}^{H}\|\hat Y_h - Y_h\|_F^2$.
For pretrained TSFMs deployed zero-shot, the inner minimization in
Eq.~\eqref{eq:erm} is replaced by direct evaluation at a fixed parameter
$\hat\theta_{\mathrm{pre}}$; the theoretical implications of this replacement
are the subject of Section~\ref{sec:pretraining-prior}.

\subsection{Pretraining as an Inductive Prior}
\label{sec:pretraining-prior}

A useful theoretical lens for interpreting the benchmark is to view pretraining
as a data-dependent prior over predictor space. Let
$\mathcal{F}_\Theta = \{f_\theta : \theta \in \Theta\}$ be a parametric family
of forecasters and let $\mathcal{D}_{\mathrm{pre}}$ denote a generic pretraining
corpus (in our setting, the diverse multi-domain corpora used to train TimeGPT,
TimesFM-2.5, Moirai-2.0, Chronos, and Chronos-2). A pretrained TSFM corresponds
to a parameter setting $\hat\theta_{\mathrm{pre}}$ obtained by minimizing an
empirical risk on $\mathcal{D}_{\mathrm{pre}}$, possibly with regularization. In
a Bayesian reading this defines an informative prior
\begin{equation}
\pi_{\mathrm{pre}}(\theta) \propto
\pi_0(\theta)\,\exp\!\big(-n_{\mathrm{pre}}\,\widehat{\mathcal{R}}_{\mathrm{pre}}(\theta)\big);
\end{equation}
in a frequentist reading it identifies a low-dimensional submanifold of $\Theta$
around which the model class's effective complexity is much smaller than the
nominal parameter count. Either reading produces the same operational
prediction: the variance of the estimator at $\hat\theta_{\mathrm{pre}}$ is
controlled by the prior dispersion rather than by the unrestricted capacity of
the architecture.

\paragraph{A small-sample generalization argument.} The relevance for our
small-data per-asset regime is summarized by a generalization bound of the form
\begin{equation}
\mathcal{R}(\hat f) - \min_{f \in \mathcal{F}_\Theta} \mathcal{R}(f)
\;\le\; \widetilde{O}\!\left(\sqrt{\frac{d_{\mathrm{eff}}(\mathcal{F}_\Theta)}{n_{\mathrm{loc}}}}\right),
\label{eq:gen-bound}
\end{equation}
where $n_{\mathrm{loc}}$ is the effective number of local samples available for
estimating $\hat f$ and $d_{\mathrm{eff}}$ is an effective complexity term
(Rademacher, VC, or PAC-Bayes complexity, depending on the framing). For
locally trained baselines $d_{\mathrm{eff}}$ reflects the full capacity of the
architecture; for a pretrained TSFM evaluated zero-shot, $d_{\mathrm{eff}}$
is governed by the local dispersion of the pretraining prior $\pi_{\mathrm{pre}}$ around
$\hat\theta_{\mathrm{pre}}$, rather than by the full nominal parameter count.
This dispersion may be much smaller than the unrestricted complexity of the
architecture, but it remains an assumption of the inductive-prior argument rather
than a quantity estimated in this benchmark. When $n_{\mathrm{loc}}$ is small relative to the unrestricted
$d_{\mathrm{eff}}$, Eq.~\eqref{eq:gen-bound} predicts the pretrained estimator
to be favored not because it ``knows finance'' but because its prior dispersion
is a tighter fit for the available local sample size.

\paragraph{What this argument does \emph{not} show.} Two negative observations
follow directly. First, dominance of pretrained TSFMs in this regime does not
isolate architectural quality: $d_{\mathrm{eff}}$ is conflated with the
pretraining prior $\pi_{\mathrm{pre}}$, so an architectural verdict over
locally trained baselines is unwarranted from rank alone. Second, the same
argument predicts that a \emph{less} financially relevant prior may still
outperform a high-capacity locally-trained model when $n_{\mathrm{loc}}$ is
small enough; the benchmark therefore cannot establish that pretraining has
captured finance-specific structure. The contemporaneous evidence of
\citet{rahimikia2025revisiting}---that finance-native pre-training delivers
substantial gains once $n_{\mathrm{loc}}$ scales to billions of observations---is
exactly the regime in which the prior contribution to Eq.~\eqref{eq:gen-bound}
becomes negligible and domain alignment of $\mathcal{D}_{\mathrm{pre}}$ becomes
the binding constraint.

\subsection{A PAC-Bayes Bound for Zero-Shot Transfer}
\label{sec:pac-bayes}

A more refined generalization argument tailored to the pretraining setting is
the PAC-Bayes bound \citep{mcallester1999pac, catoni2007pac}. Let
$\pi_{\mathrm{pre}}$ be the prior over $\Theta$ defined in
Section~\ref{sec:pretraining-prior}, let $Q$ be a posterior over $\Theta$
inferred from local financial data $\mathcal{D}_{\mathrm{loc}}$ of size
$n_{\mathrm{loc}}$, and let $\hat f$ be the corresponding randomized
predictor. For a loss $\ell \in [0, B]$ and confidence $\delta \in (0, 1)$,
with probability at least $1 - \delta$ over the draw of
$\mathcal{D}_{\mathrm{loc}}$,
\begin{equation}
\mathbb{E}_{f \sim Q}\!\big[\mathcal{R}(f)\big]
\;\le\;
\mathbb{E}_{f \sim Q}\!\big[\widehat{\mathcal{R}}_{n}(f)\big]
\;+\; \sqrt{\frac{B^{2}}{2\,n_{\mathrm{loc}}}\!\left(\mathrm{KL}\!\big(Q \,\big\|\, \pi_{\mathrm{pre}}\big) + \log \tfrac{1}{\delta}\right)}.
\label{eq:pac-bayes}
\end{equation}

Two regimes have qualitatively different interpretations of
Eq.~\eqref{eq:pac-bayes}. \emph{Pure zero-shot deployment} is best read as the
small-variance limit of a posterior concentrated near the pretrained parameter
$\hat\theta_{\mathrm{pre}}$. In a continuous parameter space, a literal Dirac
posterior $\delta_{\hat\theta_{\mathrm{pre}}}$ is generally singular with
respect to a continuous prior and can yield an infinite KL divergence, so the
zero-shot PAC-Bayes interpretation is heuristic unless one works with a
discrete parameter space or an explicitly smoothed posterior such as
$Q_{\tau}=\mathcal{N}(\hat\theta_{\mathrm{pre}},\tau^2 I)$. Under such a
smoothed approximation, the complexity term measures how concentrated the
pretraining prior is around the deployed parameter. \emph{Local fine-tuning}
constructs $Q$ from $\pi_{\mathrm{pre}}$ via gradient updates on
$\mathcal{D}_{\mathrm{loc}}$; the generalization gap then depends on how far
$Q$ is allowed to move from $\pi_{\mathrm{pre}}$ in KL divergence before the
second term dominates the empirical-risk reduction in the first.

Eq.~\eqref{eq:pac-bayes} formalizes the intuition that fine-tuning on a small
per-asset window cannot move $Q$ very far from $\pi_{\mathrm{pre}}$ without
paying an unsustainable generalization penalty: the maximum allowable KL
budget scales as $n_{\mathrm{loc}}$, so the available space of fine-tuned
posteriors collapses for small local samples. This is consistent with the
\citet{rahimikia2025revisiting} observation that fine-tuning yields only
marginal improvements over zero-shot inference: the KL constraint is binding.
Pre-training from scratch on financial data, by contrast, replaces
$\pi_{\mathrm{pre}}$ with a finance-aligned prior $\pi^{\mathrm{fin}}_{\mathrm{pre}}$,
relaxing the KL constraint by changing the reference measure. The bound
predicts the gap to close exactly where the empirical evidence reports it
closing.

\subsection{Information-Theoretic Predictability Bounds}
\label{sec:info-theoretic}

The variance bound of Eq.~\eqref{eq:predictability-bound} has an
information-theoretic counterpart that does not require finite second moments
and that exposes the small budget of bits available to any predictor on daily
liquid-equity returns. Let $I(r_{t+h}; \mathcal{F}_t)$ denote the (Shannon)
mutual information between the future return and the past $\sigma$-field, and
let $h(r_{t+h})$ denote the differential entropy of $r_{t+h}$. The classical
rate--distortion lower bound on minimum mean-squared error
\citep{coverthomas2006elements} gives
\begin{equation}
\mathrm{MSE}^\star \;\ge\; \frac{1}{2\pi e}\,
\exp\!\big(\,2\big(h(r_{t+h}) - I(r_{t+h};\mathcal{F}_t)\big)\big),
\label{eq:rate-distortion-bound}
\end{equation}
with equality under jointly Gaussian conditioning. Specializing to the
joint-Gaussian case yields the closed-form ceiling
\begin{equation}
1 - R^2_{\max} \;=\; \exp\!\big(-2\,I(r_{t+h};\mathcal{F}_t)\big),
\label{eq:gaussian-r2-information}
\end{equation}
so an out-of-sample $R^2_{\max}$ of $1\%$ corresponds to mutual information of
roughly $0.005$ nats per forecast---a small fraction of one bit. This is a
quantitative restatement of the difficulty of return forecasting: the past
simply does not carry many bits of information about the future, and any
sufficiently expressive architecture that can recover those bits will perform
comparably. Differences between models in our benchmark are competing for a
small information budget rather than for architectural superiority on a
high-information task.

Two refinements bear on the empirical results. First, returns are heavy-tailed,
so the Gaussian closed form Eq.~\eqref{eq:gaussian-r2-information} understates
the achievable MSE reduction relative to the general
Eq.~\eqref{eq:rate-distortion-bound}: a forecaster that recovers non-Gaussian
conditional structure---through quantized likelihoods (Chronos, Chronos-2) or
quantile heads (Moirai-2.0)---can in principle exceed the Gaussian ceiling
without violating the information bound. Second, the conditional information
$I(r_{t+h};\mathcal{F}_t)$ is itself non-stationary across regimes: returns
during volatility shocks plausibly carry more bits about the immediate future
than returns during calm periods, so the rolling-origin protocol used in
Section~\ref{sec:metrics} probes a moving average over heterogeneous
information levels. The cautious interpretation of skill scores in
Section~\ref{sec:results} is partly grounded in this regime mixing.

\subsection{Causal Capacity and Maximum Trading Utility}
\label{sec:causal-capacity}

The information bound of Section~\ref{sec:info-theoretic} sharpens into a
statement about achievable economic utility. Suppose an investor with log
utility chooses positions $\pi_t$ in the asset and a risk-free instrument
based on $\mathcal{F}_t$. The Kelly--Cover growth-optimal strategy
\citep{kelly1956new, coverthomas2006elements} maximizes the conditional
expected log-return,
\begin{equation}
\pi^\star_{t} \;=\; \arg\max_{\pi}\; \mathbb{E}\!\left[
\log\!\big(1 + \pi(r_{t+h} - r_{f})\big) \,\big|\, \mathcal{F}_{t}\right],
\end{equation}
and its asymptotic almost-sure growth rate $W^\star$ is the maximum log-wealth
achievable by any $\mathcal{F}_t$-adapted strategy. \citet{algoet1988asymptotic}
show $W^\star$ is bounded above by an information-theoretic capacity:
\begin{equation}
W^\star \;\le\; \mathcal{C}\!\big(\mathcal{F}_{t} \to r_{t+h}\big)
\;:=\; \limsup_{n \to \infty}\; \frac{1}{n}\, I\!\big(\mathcal{F}_{t}^{n} \to r_{t+h}^{n}\big),
\label{eq:causal-capacity}
\end{equation}
where the right-hand side is the directed (causal) information from the
conditioning sequence to the return sequence in the sense of
\citet{kramer1998directed, permuter2011interpretations}, and reduces to the
ordinary mutual information $I(r_{t+h};\mathcal{F}_{t})$ when $\mathcal{F}_{t}$
contains no past values of $r$.

Eq.~\eqref{eq:causal-capacity} ties the small-information regime of
Section~\ref{sec:info-theoretic} directly to a small economic edge: if the
Shannon mutual information $I(r_{t+h};\mathcal{F}_{t})$ is on the order of
$0.005$ nats, the directed-information ceiling $\mathcal{C}$ is bounded by the
same order, and the maximum log-growth attainable by any forecaster--strategy
pair before transaction costs is correspondingly small. Statistical-significance
claims for one model over another do not, by themselves, translate into
proportionally large trading edges, because both sides of any such comparison
are bounded by the same capacity. The MAE skill scores of order $10^{-3}$
reported in Section~\ref{sec:results} are consistent with what
Eq.~\eqref{eq:causal-capacity} permits; they do not prove an exploitable
trading advantage, and the discussion of investment-utility limitations in
Section~\ref{sec:discussion} uses this argument as theoretical backup.

\subsection{Operator-Theoretic View of Attention}
\label{sec:operator-theoretic}

Self-attention---the central computational primitive of TimeGPT, TimesFM-2.5,
Moirai-2.0, Chronos, Chronos-2, PatchTST, and iTransformer---admits a clean
reading as a kernel integral operator on a space of token-valued functions.
For a sequence of $N$ tokens with embeddings $\{x_i\}_{i=1}^{N} \subset
\mathbb{R}^{d}$ and learnable projections
$W_Q, W_K, W_V \in \mathbb{R}^{d_k \times d}$, scaled-dot-product attention
produces
\begin{equation}
(\mathcal{A} x)_i \;=\; \sum_{j=1}^{N} \kappa(x_i, x_j)\, W_V x_j,
\qquad
\kappa(x_i, x_j) \;=\; \frac{\exp\!\big(\langle W_Q x_i, W_K x_j\rangle / \sqrt{d_k}\big)}
{\sum_{\ell=1}^{N} \exp\!\big(\langle W_Q x_i, W_K x_\ell\rangle / \sqrt{d_k}\big)}.
\end{equation}
In the continuum limit $N \to \infty$ with $\{x_i\}$ replaced by a continuous
field $f : \mathcal{X} \to \mathbb{R}^d$ and the empirical sum by an integral
against an underlying measure $\mu$ on the index set,
\begin{equation}
(\mathcal{A} f)(x) \;=\; \int_{\mathcal{X}} \kappa(x, y)\, (W_V f)(y)\, \mu(dy),
\end{equation}
whose normalized kernel component is row-stochastic. After the value projection,
residual connection, normalization, and feed-forward layers of an actual
transformer block, the full block is not itself a Markov chain; the Markov-kernel
language below is therefore an analogy for the attention weights rather than an
exact probabilistic model of the entire network. A transformer block is then the residual map
$f \mapsto f + \sigma(\mathcal{A} f + b)$ for a coordinate-wise nonlinearity
$\sigma$, and a stack of $L$ such blocks defines a discrete-time flow
\begin{equation}
f_{\ell+1} \;=\; f_\ell + \sigma(\mathcal{A}_\ell f_\ell + b_\ell),
\qquad \ell = 0, 1, \ldots, L-1,
\end{equation}
which can be interpreted heuristically as an explicit Euler discretization of a
non-autonomous flow $\dot f = \sigma(\mathcal{A}_t f + b_t)$ on a separable
Hilbert space.

Two consequences bear on the benchmark, provided this operator view is kept
heuristic. First, spectral properties of the input-dependent attention kernels
can influence long-range mixing: wider gaps in the normalized attention kernel
correspond to faster token-to-token averaging, while near-block-diagonal patterns
slow information flow across regions of the sequence. Second, the depth-versus-width tradeoff acquires a
numerical-analysis interpretation as a stability-versus-resolution tradeoff in
the underlying ODE: too few blocks under-resolve the flow; too many push the
explicit-Euler discretization toward instability unless residual scaling is
adjusted accordingly. PatchTST, iTransformer, and the foundation-model
backbones differ primarily in the choice of $\mathcal{A}$---along the time
axis (PatchTST, decoder-only TSFMs) versus across variates (iTransformer)---and
in the discretization step size implicitly set by residual scaling. None of
these design choices is uniformly optimal, which is consistent with the
no-universal-winner pattern of asset-level results in
Section~\ref{sec:results}.

\subsection{Spectral Gap and Mixing Time of Attention}
\label{sec:spectral-gap}

The Markov-kernel language of Section~\ref{sec:operator-theoretic} should be read
as a diagnostic analogy, not as a theorem about the fixed behaviour of the
models evaluated here. Actual transformer attention kernels vary by input,
layer, and head; they are generally non-reversible; and the benchmark does not
estimate their spectra. With that caveat, a single attention head for one input
sequence defines a row-stochastic matrix $K \in \mathbb{R}^{N \times N}$ over
tokens. If that matrix is irreducible and aperiodic, it has at least one
stationary distribution $\mu$, and in the reversible idealization its spectral
gap controls the mixing bound
\begin{equation}
t_{\mathrm{mix}}(\varepsilon)
\;\le\; \frac{1}{1 - \lambda_{2}}\,
\log\!\frac{1}{\mu_{\min}\,\varepsilon},
\label{eq:mixing-time}
\end{equation}
where $\lambda_2$ is the second-largest eigenvalue in modulus. For non-reversible
attention matrices, analogous statements use singular values, pseudo-spectral
quantities, or conductance bounds rather than this simple eigenvalue expression.

Under the reversible idealization, Cheeger's inequality
\citep{lawlersokal1988bounds} relates the gap to the conductance of the kernel,
\begin{equation}
\Phi(K) \;:=\; \min_{S : \mu(S) \le 1/2}\;
\frac{\sum_{x \in S,\, y \notin S} \mu(x)\, K(x, y)}{\mu(S)},
\qquad
\frac{\Phi(K)^{2}}{2} \;\le\; 1 - \lambda_{2} \;\le\; 2\, \Phi(K).
\label{eq:cheeger}
\end{equation}
A near-block-diagonal attention pattern has small conductance and therefore slow
cross-block information flow in this simplified model. Conversely, dense global
attention permits faster token mixing. These observations are useful for
interpreting why deep decoder-only TSFMs may carry long-context information more
comfortably than shallow locally trained models, but the present benchmark does
not measure attention gaps and therefore cannot use them as causal evidence for
any particular model ranking. The empirical conclusion remains narrower:
architectural depth and long-context machinery may matter less in daily return
forecasting than in high-information domains because the available conditional
information in the return series is itself small.

\subsection{Signatures and the Theory of Patching}
\label{sec:signatures}

The patching mechanism shared by PatchTST, TimesFM-2.5, and Moirai-2.0 admits a
rough-paths interpretation that clarifies what information patches do and do
not encode. Given a continuous path $X : [0, T] \to \mathbb{R}^d$, the
\emph{signature} of $X$ is the formal series of iterated integrals
\begin{equation}
S_{[s,t]}(X) \;=\; \Big( 1,\; \int_{s}^{t} dX_{u_1},\;
\int_{s < u_1 < u_2 < t} dX_{u_1} \otimes dX_{u_2},\;\ldots \Big),
\end{equation}
and the celebrated theorem of \citet{chen1958integration} together with the
extension of \citet{lyons1998differential} establishes that the signature
characterizes $X$ up to tree-like equivalence and that any continuous
path-dependent functional can be approximated arbitrarily well by a linear
functional of a sufficiently high-order signature truncation
\citep{chevyrev2016primer}.

A patch of length $P$ on a sampled path is not automatically a path
signature. It is a finite vector of local observations that can be interpreted
as a learned summary of path increments and within-patch shape. A linear patch
projection can, in principle, approximate low-order functionals of those local
increments, including quantities reminiscent of truncated signatures, but the
model is not constrained to compute signature coordinates and the benchmark does
not test whether it does so. Likewise, channel independence and instance
normalization help stabilize scale and distribution shift, but they do not by
themselves enforce the reparameterization invariance associated with rough-path
signatures. From this viewpoint, patch length $P$ and embedding dimension $D$
control the amount of local path information made available to the transformer,
not a literal signature truncation level.

Two practical implications follow at the level of representation design. First,
very short patches expose fine local variation but increase the number of tokens
and force the transformer stack to aggregate local structure through attention.
Second, long patches reduce token count but compress more observations into a
single embedding, which can discard within-patch detail when the embedding
capacity is limited. The standard PatchTST default of $P = 16$ with $D = 128$ is
one reasonable point in this compression--resolution tradeoff; we use the
canonical configuration without tuning, which is consistent with our as-used
framing in Section~\ref{sec:methodology} but should be treated as a design
choice rather than an optimized setting.

\subsection{Information Geometry of Pretrained Forecasters}
\label{sec:information-geometry}

The parameter manifold of any probabilistic forecaster carries a natural
Riemannian metric, the Fisher information metric
\begin{equation}
g_{ij}(\theta) \;=\; \mathbb{E}_{p_\theta}\!\left[
\frac{\partial \log p_\theta(y \mid x)}{\partial \theta^i}\,
\frac{\partial \log p_\theta(y \mid x)}{\partial \theta^j}\right],
\label{eq:fisher-metric}
\end{equation}
which equips $(\Theta, g)$ with a Riemannian structure and induces the
natural-gradient direction $g^{-1}\nabla \mathcal{L}$ that is invariant under
smooth reparameterization \citep{amari1998natural}. Two specializations are
relevant here.

\paragraph{Categorical likelihoods (Chronos, Chronos-2).} For the $B$-bin
quantized likelihood $p_\theta(z = i \mid x) = \pi_i(\theta; x)$, the Fisher
metric is the multinomial information matrix
\begin{equation}
g_{ij}(\theta) \;=\; \sum_{k=1}^{B}
\frac{1}{\pi_k} \frac{\partial \pi_k}{\partial \theta^i}
\frac{\partial \pi_k}{\partial \theta^j},
\end{equation}
whose curvature concentrates near the boundary of the probability simplex. In
practical terms, models that place near-deterministic mass on a single
quantile bin (the regime relevant for tail forecasts on volatile days) sit in
high-curvature regions of $(\Theta, g)$ where small parameter perturbations
produce large changes in conditional distribution.

\paragraph{Pinball-based quantile heads (Moirai-2.0).} The pinball loss is
not a likelihood, but its second-order behaviour around the conditional
$q$-quantile defines an analogous geometry in which the local metric is
concentrated near the conditional median ($q = 0.5$) and degrades smoothly
toward the tails. The natural-gradient analogue here is the pre-conditioner
defined by the second derivative of the dual cumulant of the pinball loss; we
do not compute it explicitly, but it explains why quantile-head forecasters are
typically well-behaved for central quantiles and progressively less
well-conditioned for extreme ones.

\paragraph{Connection to the inductive-prior view.} The pretraining prior
$\pi_{\mathrm{pre}}(\theta)$ has effective support concentrated on a
low-dimensional submanifold of $(\Theta, g)$, and the Fisher-geodesic distance
from $\hat\theta_{\mathrm{pre}}$ to its nearest finance-optimal parameter
setting controls how much benefit a small amount of fine-tuning could in
principle deliver. The empirical observation of \citet{rahimikia2025revisiting}
that fine-tuning yields only marginal improvements over zero-shot, while
pre-training from scratch on financial data yields substantial gains, is
consistent with the corresponding Fisher-distance picture: the generic
pretrained submanifold is far enough from the finance-optimal submanifold,
relative to the Fisher length scale, that local moves do not bridge it. We do
not estimate Fisher distances for the systems in this benchmark, but the
conceptual implication is direct: genuinely improving zero-shot transfer to
finance requires either extending $\mathcal{D}_{\mathrm{pre}}$ to include
financial data or re-pretraining from scratch on a finance-aligned corpus.
Both operations modify $\pi_{\mathrm{pre}}$, not the local fine-tuning step.

\subsection{Distribution Geometry of Forecast Scores}
\label{sec:wasserstein-geometry}

The quantile-score discussion for Moirai-2.0 is best expressed as an integrated
quantile-loss or CRPS geometry, with Wasserstein distance used only as a related
but distinct comparison metric. For $p \ge 1$, the $p$-Wasserstein distance
between probability measures $F, G$ on $\mathbb{R}$ with finite $p$-th moments
has the one-dimensional quantile form
\begin{equation}
W_{p}(F, G)^{p} \;=\; \int_{0}^{1} \left| F^{-1}(q) - G^{-1}(q) \right|^{p} dq.
\label{eq:wp-quantile}
\end{equation}
When $G=\delta_y$ and $p=1$, this becomes
$W_1(F,\delta_y)=\mathbb{E}|X-y|$. CRPS instead subtracts the dispersion term in
Eq.~\eqref{eq:crps-energy}; it rewards both calibration and sharpness and is a
strictly proper scoring rule for the full predictive distribution. Thus the
benchmark should not describe quantile-head models as minimizing Wasserstein
score in the literal sense. The accurate statement is that finite quantile
training approximates an integrated pinball-loss objective, which is equivalent
to CRPS up to the normalization convention in Eq.~\eqref{eq:crps}.

Wasserstein geometry remains useful as an auxiliary language for comparing
forecast distributions. For example, in Wasserstein-$2$ space,
$\mathcal{W}_{2}(\mathbb{R})$ \citep{otto2001geometry,villani2003topics},
geodesics interpolate predictive distributions
linearly in quantile space:
\begin{equation}
F_{t}^{-1}(q) \;=\; (1-t)\, F_{0}^{-1}(q) + t\, F_{1}^{-1}(q),
\qquad t \in [0, 1].
\end{equation}
For Gaussian forecast distributions
$F = \mathcal{N}(\mu_{F}, \sigma_{F}^{2})$ and
$G = \mathcal{N}(\mu_{G}, \sigma_{G}^{2})$,
Eq.~\eqref{eq:wp-quantile} specializes to
\begin{equation}
W_{2}(F, G)^{2} \;=
(\mu_{F} - \mu_{G})^{2} + (\sigma_{F} - \sigma_{G})^{2},
\label{eq:w2-gaussian}
\end{equation}
which separates location error from scale error. This decomposition is helpful
for intuition: point-forecast baselines mainly target the location component,
whereas probabilistic or quantile models can represent both location and
conditional dispersion. It does not imply that the MAE rankings in this paper are
Wasserstein rankings, because the reported benchmark metric is point MAE.

This distinction matters for interpretation. Moirai-2.0's wins on GOOG and on the
AMZN log-return task may be consistent with a probabilistic model benefiting from
conditional-dispersion information, but the evidence is suggestive rather than
conclusive. Other features of the evaluation window---idiosyncratic events,
volatility clustering, and asset-specific valuation dynamics---may also explain
these results. A direct test of the geometric hypothesis would require evaluating
probabilistic scores such as CRPS, calibration curves, and quantile coverage,
none of which is part of the present MAE-only benchmark.

\subsection{Cross-Modality Reprogramming (Time-LLM)}

The cross-modality reprogramming approach pioneered by Time-LLM
\citep{jin2024timellmtimeseriesforecasting} leverages the high-dimensional latent
representations of frozen LLMs, such as Llama
\citep{touvron2023llamaopenefficientfoundation} or GPT-2
\citep{radford2019languagemodelsareunsupervised}, and adapts them for temporal
sequence analysis. The adaptation typically consists of three stages:
\begin{enumerate}[leftmargin=1.5em]
    \item \textbf{Input transformation and patching.} Following the channel-independence
    strategy \citep{nie2023timeseriesworth64}, each multivariate series is treated as
    $N$ independent univariate sequences $X^{(i)} \in \mathbb{R}^{1 \times T}$. These
    sequences undergo normalization (typically RevIN) and are segmented into
    overlapping patches to preserve local temporal dependencies and reduce input
    dimensionality.
    \item \textbf{Reprogramming with text prototypes.} To bridge continuous numerical
    data and the discrete token space of the LLM, a set of learned text prototypes
    aligns the source temporal modality with the target linguistic modality, so that
    the frozen backbone can ``interpret'' time-series patterns as semantic sequences.
    \item \textbf{Prompt-as-Prefix and frozen backbone.} Task-specific instructions in
    natural language (e.g.\ ``forecast the next 20 days of stock returns'') are
    prepended to the reprogrammed embeddings and processed by the pretrained
    transformer layers, which are kept frozen.
\end{enumerate}
Only the lightweight input transformation layers, the text prototypes, and the
output projection head are updated during training, which makes this approach
parameter-efficient but keeps Time-LLM conceptually adjacent to, rather than part
of, the empirical benchmark below.

\begin{figure}[H]
    \centering
    \includegraphics[width=0.9\linewidth]{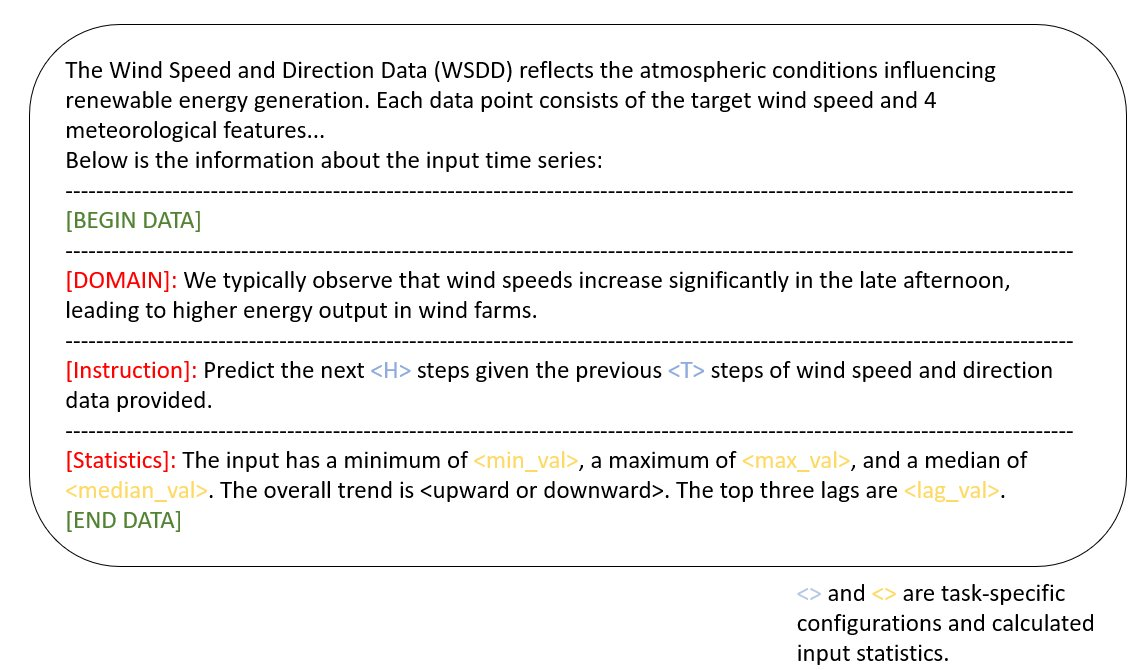}
    \caption{A concrete Prompt-as-Prefix example that conditions a frozen LLM on
    task context, instructions, and summary statistics of the input series \citep{jin2024timellmtimeseriesforecasting}.}
    \label{fig:Prompt_example}
\end{figure}

\subsection{TimeGPT and TimeGPT Long Horizon}

TimeGPT \citep{garza2024timegpt1} is a foundation model designed for time-series
forecasting that leverages transfer learning to perform zero-shot inference. It
provides a mapping $f_{\theta}: \mathcal{X} \mapsto \mathcal{Y}$ with
$\mathcal{X} = \{y_{[0:t]}, x_{[0:t+h]}\}$ (historical values and exogenous
covariates) and $\mathcal{Y} = \{y_{[t+1:t+h]}\}$ for forecast horizon $h$:
\begin{equation}
\mathbb{P}(y_{[t+1:t+h]} \mid y_{[0:t]}, x_{[0:t+h]}) = f_{\theta}(y_{[0:t]}, x_{[0:t+h]}).
\end{equation}
The architecture is a Transformer encoder--decoder with self-attention and local
positional encoding. TimeGPT was trained on over 100 billion data points spanning
diverse frequencies and domains. Reported performance is typically scale-independent,
using relative Mean Absolute Error (rMAE) and relative Root Mean Square Error
(rRMSE) against a Seasonal Naive baseline:
\begin{equation}
\mathrm{rMAE} = \frac{\sum_{i=1}^{n} \sum_{t=1}^{h} |y_{i,t} - \hat{y}_{i,t}|}{\sum_{i=1}^{n} \sum_{t=1}^{h} |y_{i,t} - \hat{y}_{i,t}^{\mathrm{base}}|},
\qquad
\mathrm{rRMSE} = \sqrt{ \frac{\sum_{i=1}^{n} \sum_{t=1}^{h} (y_{i,t} - \hat{y}_{i,t})^2}{\sum_{i=1}^{n} \sum_{t=1}^{h} (y_{i,t} - \hat{y}_{i,t}^{\mathrm{base}})^2} }.
\end{equation}
In TimeGPT Long Horizon (TimeGPT-LH) scenarios, uncertainty is quantified through
a non-parametric conformal prediction framework. Rolling forecasts on the latest
available data estimate errors and yield prediction intervals without strict
distributional assumptions, which is useful for the heavy-tailed behaviour common
in financial datasets.

\subsection{TimesFM-2.5: Decoder-Only Foundation Model}

TimesFM \citep{das2024decoderonlyfoundationmodeltimeseries} is a decoder-only
transformer architecture designed for high-accuracy zero-shot forecasts. It is
pretrained on approximately 100 billion time points, including real-world data
from Google Trends and Wiki Pageviews, and synthetic data from ARMA processes and
seasonal-trend decompositions.

The model treats time-series segments as tokens via a patching mechanism. Given an
input sequence $y_{1:L}$, the data are divided into contiguous, non-overlapping
patches of length $p$. The $j$-th patch $y_j = y_{p(j-1)+1:pj}$ is processed by an
input residual block to produce
\begin{equation}
t_j = \mathrm{InputResidualBlock}(\bar{y}_j \odot (1 - \bar{m}_j)) + PE_j,
\end{equation}
where $\bar{m}_j$ is a binary padding mask and $PE_j$ is the $j$-th positional
encoding. The output token is produced by a causal stack,
\begin{equation}
o_j = \mathrm{StackedTransformer}((t_1, \dot{m}_1), \dots, (t_j, \dot{m}_j)),
\end{equation}
and the subsequent $h$ time points after the $j$-th patch are generated by an
output residual block:
\begin{equation}
\hat{y}_{pj+1:pj+h} = \mathrm{OutputResidualBlock}(o_j).
\end{equation}
Training minimizes the mean squared error across all predicted patches in a
mini-batch:
\begin{equation}
\mathcal{L} = \frac{1}{N} \sum_{j=1}^{N} \mathrm{MSE}(\hat{y}_{pj+1:pj+h}, y_{pj+1:pj+h}).
\end{equation}
A distinctive feature is that the output patch length $h$ can exceed the input
patch length $p$, reducing the number of autoregressive steps required at inference
for long-horizon forecasting.

\subsection{Moirai-2.0: Decoder-Only Universal Transformer}

Moirai-2.0 \citep{liu2026moirai20timeseries} transitions from a masked-encoder to a
decoder-only architecture. It is trained on roughly 36 million time series
(approximately 295 billion observations) and is optimized for both probabilistic
accuracy and inference efficiency.

Univariate series are partitioned into $T$ contiguous, non-overlapping patches. To
avoid future information leakage, instance-normalization statistics are computed
solely from the first 30\% of the sequence. Each patch $\hat{x}_i$ is mapped to a
$d$-dimensional embedding via a residual block with SiLU activation:
\begin{equation}
t_{i} = \mathrm{SiLU}(W(\hat{x}_{i}) + b) + \hat{x}_{i} \in \mathbb{R}^{d}.
\end{equation}
A core design choice is the replacement of mixture-distribution outputs with
\emph{quantile forecasting}. The model predicts $n_q = 9$ quantile levels
$Q = \{0.1, 0.2, \dots, 0.9\}$ by minimizing the pinball loss
\begin{equation}
l_{q}(y_{t}, \hat{y}_{t}^{(q)}) =
\begin{cases}
q(y_{t} - \hat{y}_{t}^{(q)}) & \text{if } y_{t} \geq \hat{y}_{t}^{(q)}, \\
(1-q)(\hat{y}_{t}^{(q)} - y_{t}) & \text{if } \hat{y}_{t}^{(q)} > y_{t}.
\end{cases}
\end{equation}
The total objective averages this loss across quantiles and time steps $H$ in $K$
predicted patches:
\begin{equation}
\mathcal{L}_{Q} = \frac{1}{H|Q|} \sum_{t=1}^{H} \sum_{q \in Q} \left[ q \max(y_{t} - \hat{y}_{t}^{(q)}, 0) + (1-q) \max(\hat{y}_{t}^{(q)} - y_{t}, 0) \right].
\end{equation}
For long horizons, Moirai-2.0 generates forecasts for $n_{\mathrm{token}}$ future
patches simultaneously to reduce error accumulation. At inference, an autoregressive
multi-quantile decoding strategy---conceptually similar to a beam search of depth
two---expands and collapses forecast candidates back into the target quantile set,
preserving uncertainty estimates.

\paragraph{Pinball loss as a strictly proper scoring rule.} The pinball loss
$\ell_q(y, \hat y^{(q)}) = \max\{q(y - \hat y^{(q)}),\, (q-1)(y - \hat y^{(q)})\}$
is, up to affine equivalence, the unique loss function whose population
minimizer over $\hat y^{(q)}$ is the conditional $q$-quantile of $Y$
\citep{koenker2005quantile}. Pointwise consistency therefore implies that any
forecaster trained to minimize pinball loss at a level $q$ produces, at the
population minimum, the conditional $q$-quantile of the response.

When predictions $\hat y^{(q_1)} \le \cdots \le \hat y^{(q_K)}$ are produced
for ordered levels $q_1 < \cdots < q_K$, the average pinball loss
\begin{equation}
\bar \ell\!\big(y, \hat y^{(\cdot)}\big)
\;=\; \frac{1}{K} \sum_{k=1}^{K} \ell_{q_k}\!\big(y, \hat y^{(q_k)}\big)
\end{equation}
is a finite-sample approximation to the continuous ranked probability score
(CRPS),
\begin{equation}
\mathrm{CRPS}(F, y)
\;=\; \int_{\mathbb{R}} \big(F(z) - \mathbf{1}\{z \ge y\}\big)^{2} dz
\;=\; 2 \int_{0}^{1} \ell_{q}\!\big(y, F^{-1}(q)\big)\, dq,
\label{eq:crps}
\end{equation}
which is a strictly proper scoring rule for one-dimensional predictive
distributions \citep{gneiting2007strictly}. CRPS is closely related to the
integrated quantile loss, but it is not equal to the Wasserstein-$1$ distance
from $F$ to a point mass. If $X,X' \sim F$ independently, then
\begin{equation}
\mathrm{CRPS}(F,y)
\;=\; \mathbb{E}|X-y| - \frac{1}{2}\mathbb{E}|X-X'|,
\qquad
W_1(F,\delta_y) \;=\; \mathbb{E}|X-y|.
\label{eq:crps-energy}
\end{equation}
The second term in Eq.~\eqref{eq:crps-energy} is the sharpness correction that
makes CRPS a proper distributional score rather than a pure distance to the
realized observation. Moirai-2.0's quantile head is therefore optimizing a
discrete approximation to an integrated quantile/CRPS objective whose population
minimum is achieved by the conditional law of the response. The non-crossing
constraint enforced by the autoregressive multi-quantile decoder is the
operational analogue of imposing a monotone $F^{-1}$; without monotonicity, the
predicted quantiles may fail to define a valid predictive distribution.

\subsection{NBEATS}

\citet{oreshkin2020nbeatsneuralbasisexpansion} propose a basic building block with
a fork architecture. The $\ell$-th block accepts an input $\mathbf{x}_{\ell}$ and
produces two outputs: $\hat{\mathbf{x}}_{\ell}$ (the backcast) and
$\hat{\mathbf{y}}_{\ell}$ (the forward forecast).

\begin{figure}[H]
    \centering
    \caption{NBEATS architecture.}
    \includegraphics[width=\linewidth]{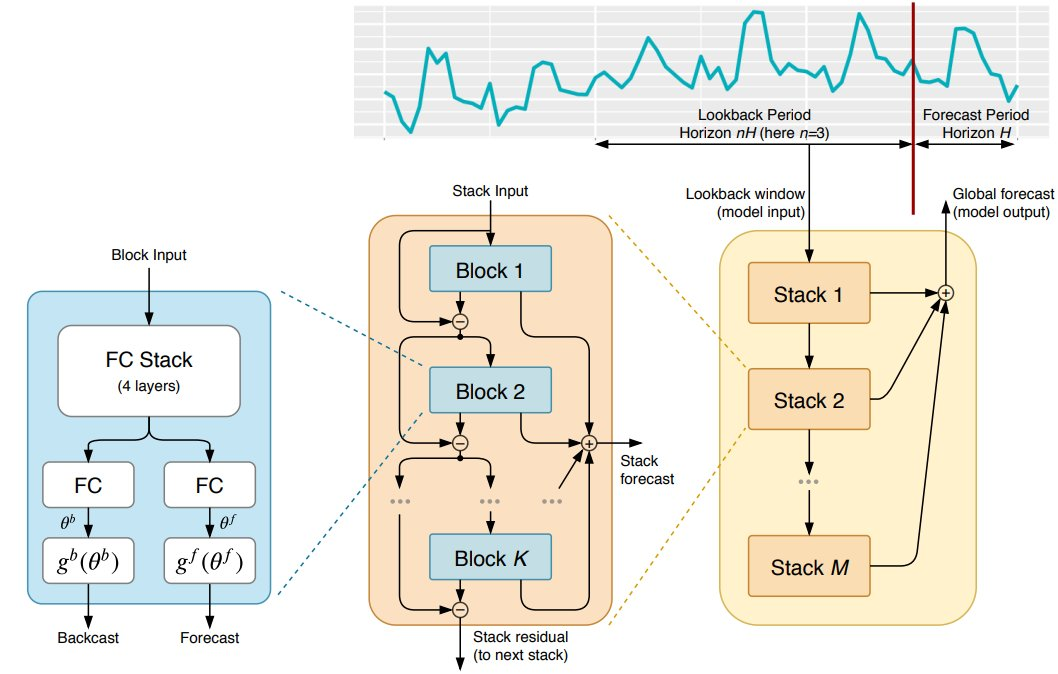}
    \caption*{Source: \citet{oreshkin2020nbeatsneuralbasisexpansion}.}
    \label{fig:nbeats}
\end{figure}

The first block in the model takes a history lookback window as its input
$\mathbf{x}_{\ell}$, typically of length $2H$ to $7H$. For subsequent blocks the
input is the residual output from preceding blocks. Each block has two parts: a
fully connected network that produces forward $\theta^{f}_{\ell}$ and backward
$\theta^{b}_{\ell}$ expansion coefficients, and basis layers $g^{f}_{\ell}$,
$g^{b}_{\ell}$ that project those coefficients onto basis functions to generate
$\hat{\mathbf{y}}_{\ell}$ and $\hat{\mathbf{x}}_{\ell}$. The first part is
described by
\begin{align*}
\mathbf{h}_{\ell,1} &= \mathrm{FC}_{\ell,1}(\mathbf{x}_{\ell}), \\
\mathbf{h}_{\ell,2} &= \mathrm{FC}_{\ell,2}(\mathbf{h}_{\ell,1}), \\
\mathbf{h}_{\ell,3} &= \mathrm{FC}_{\ell,3}(\mathbf{h}_{\ell,2}), \\
\mathbf{h}_{\ell,4} &= \mathrm{FC}_{\ell,4}(\mathbf{h}_{\ell,3}), \\
\theta^{b}_{\ell} &= \mathrm{LINEAR}^{b}_{\ell}(\mathbf{h}_{\ell,4}), \\
\theta^{f}_{\ell} &= \mathrm{LINEAR}^{f}_{\ell}(\mathbf{h}_{\ell,4}),
\end{align*}
where each FC layer applies a ReLU non-linearity. The second part maps the
expansion coefficients to outputs:
\begin{align*}
\hat{\mathbf{y}}_{\ell} &= \sum_{i=1}^{\dim(\theta^{f}_{\ell})} \theta^{f}_{\ell,i}\, \mathbf{v}^{f}_{i}, &
\hat{\mathbf{x}}_{\ell} &= \sum_{i=1}^{\dim(\theta^{b}_{\ell})} \theta^{b}_{\ell,i}\, \mathbf{v}^{b}_{i}.
\end{align*}

\subsection{NHITS}

\citet{challu2022nhitsneuralhierarchicalinterpolation} build upon NBEATS with
multi-rate sampling of the input signal and multi-scale synthesis of the forecast,
yielding a hierarchical forecast construction that reduces computational demands
while improving long-horizon accuracy.

\begin{figure}[H]
    \centering
    \caption{N-HiTS architecture.}
    \includegraphics[width=\linewidth]{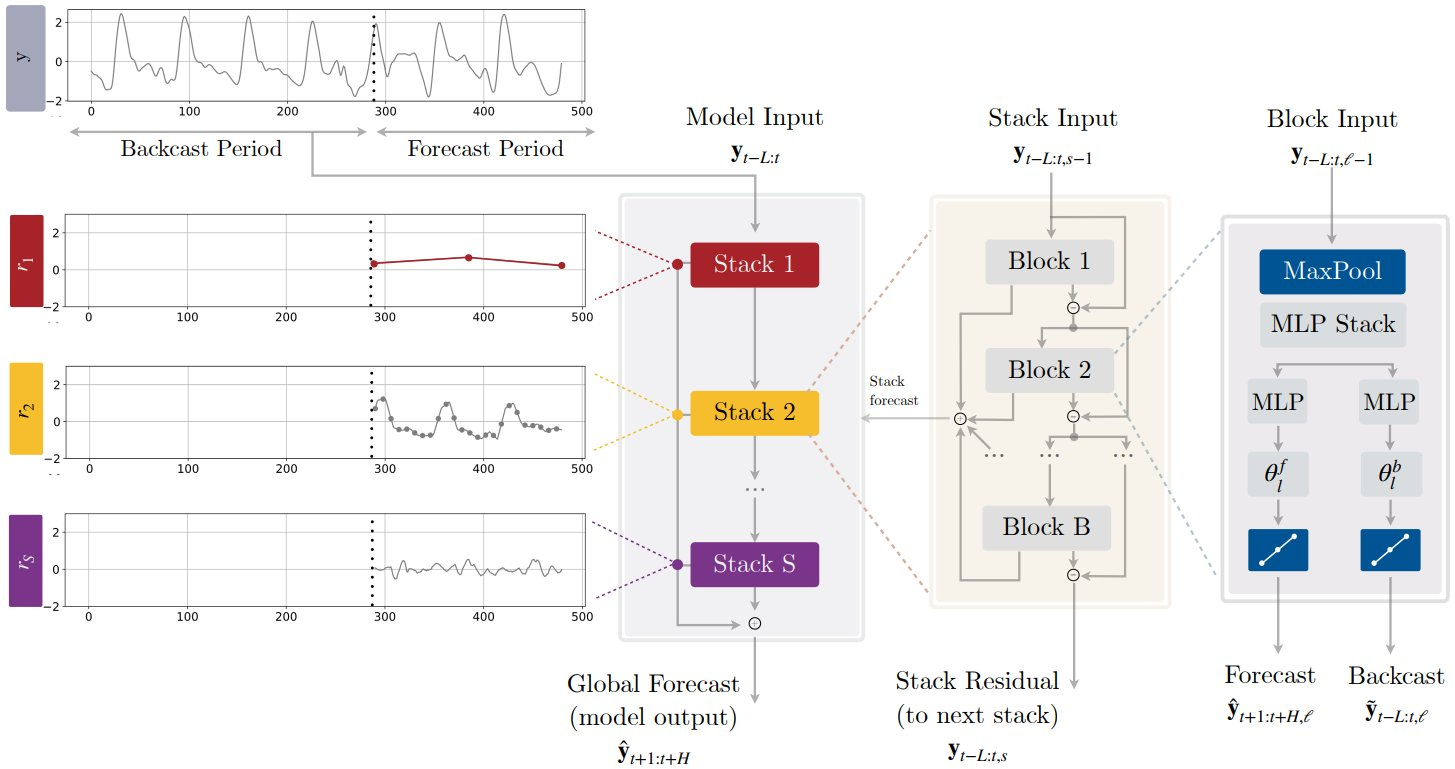}
    \caption*{Source: \citet{challu2022nhitsneuralhierarchicalinterpolation}.}
    \label{fig:nhits}
\end{figure}

Like NBEATS, NHITS performs local nonlinear projections onto basis functions across
multiple blocks, each comprising an MLP that generates backcast and forecast
coefficients. Blocks are organized into stacks, each specializing in a distinct
characteristic of the data with its own set of basis functions. A MaxPool layer is
introduced at the input of each block $\ell$,
\begin{equation}
y^{(p)}_{t-L:t,\ell} = \mathrm{MaxPool}(y_{t-L:t,\ell}, k_{\ell}),
\end{equation}
so that larger kernel sizes $k_{\ell}$ remove high-frequency components and force
the block to focus on large-scale, low-frequency content. After subsampling, each
block produces a hidden vector $h_{\ell}$ and forward and backward coefficients:
\begin{align*}
h_{\ell} &= \mathrm{MLP}_{\ell}(y^{(p)}_{t-L:t,\ell}), \\
\theta^{f}_{\ell} &= \mathrm{LINEAR}^{f}(h_{\ell}), \\
\theta^{b}_{\ell} &= \mathrm{LINEAR}^{b}(h_{\ell}).
\end{align*}
To handle long horizons, NHITS uses \emph{hierarchical interpolation}: the
dimensionality of the forward interpolation coefficients is controlled by an
expressiveness ratio $r_{\ell}$, with $|\theta|^{f}_{\ell} = \lceil r_{\ell} H \rceil$.
A temporal interpolation function $g$ recovers the full horizon, and the final
forecast sums the outputs of all blocks while backcast residuals are subtracted
from inputs of subsequent hierarchy levels.

\subsection{PatchTST}

\citet{nie2023timeseriesworth64} propose PatchTST for multivariate time-series
forecasting with a lookback window $L$, using a vanilla Transformer encoder as
backbone.

\begin{figure}[H]
    \centering
    \caption{PatchTST illustrative multivariate forecasting case study.}
    \includegraphics[width=\linewidth]{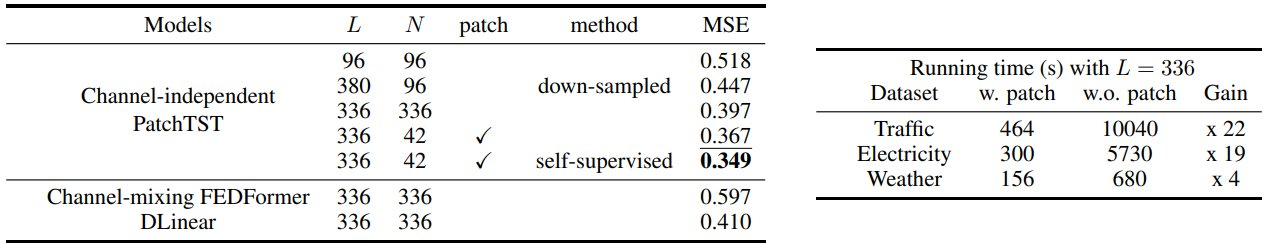}
    \caption*{Source: \citet{nie2023timeseriesworth64}.}
    \label{fig:PatchTST}
\end{figure}

A univariate series of length $L$, denoted $x^{(i)}_{1:L}$ for
$i = 1, \ldots, M$, is fed into the backbone under a channel-independence scheme
and produces prediction results
$\hat{x}^{(i)} = (\hat{x}^{(i)}_{L+1}, \ldots, \hat{x}^{(i)}_{L+T}) \in \mathbb{R}^{1 \times T}$.
Each series is partitioned into (possibly overlapping) patches of length $P$ with
stride $S$, yielding patches $x^{(i)}_p \in \mathbb{R}^{P \times N}$ where $N$ is
the number of patches; this reduces the number of input tokens from $L$ to
approximately $L/S$ and enables longer histories to be processed.

Patches are mapped to the latent dimension $D$ via a trainable linear projection
$W_{p} \in \mathbb{R}^{D \times P}$ with a learnable additive positional encoding
$W_{\mathrm{pos}} \in \mathbb{R}^{D \times N}$. Multi-head attention then operates
on these tokens:
\begin{align*}
\left( O^{(i)}_{h} \right)^{\top} &= \mathrm{Attention}(Q^{(i)}_{h}, K^{(i)}_{h}, V^{(i)}_{h}) \\
&= \mathrm{Softmax}\!\left( \frac{Q^{(i)}_{h} K^{(i)\top}_{h}}{\sqrt{d_k}} \right) V^{(i)}_{h}.
\end{align*}
The multi-head block uses BatchNorm and a feed-forward network with residual
connections, producing $z^{(i)} \in \mathbb{R}^{D \times N}$; a flatten plus linear
head produces the final prediction $\hat{x}^{(i)}$. The training loss is
\begin{equation}
\mathcal{L} = \mathbb{E}_{x}\, \frac{1}{M} \sum_{i=1}^{M} \| \hat{x}^{(i)}_{L+1:L+T} - x_{L+1:L+T} \|^{2}_{2}.
\end{equation}
An instance-normalization technique zero-means and unit-scales each series before
patching, adding the statistics back to the output to mitigate train--test
distribution shift.

\subsection{Chronos}

\citet{ansari2024chronoslearninglanguagetime} introduce Chronos, a framework that
adapts existing language-model architectures and training procedures to probabilistic
time-series forecasting. The key step is tokenizing real-valued time series via
scaling and quantization. Given $x_{1:C+H} = [x_1, \ldots, x_{C+H}]$, mean scaling
gives $\tilde{x}_i = x_i / s$ with $s = \tfrac{1}{C}\sum_{i=1}^{C}|x_i|$.
Quantization with $B$ bin centers $c_1 < \cdots < c_B$ and $B-1$ edges $b_i$ maps
scaled values to discrete tokens:
\begin{equation}
q(x) =
\begin{cases}
1 & \text{if } -\infty \leq x < b_1, \\
2 & \text{if } b_1 \leq x < b_2, \\
\;\vdots \\
B & \text{if } b_{B-1} \leq x < \infty,
\end{cases}
\qquad d(j) = c_j.
\end{equation}
Chronos uses both encoder--decoder (T5) and decoder-only (GPT-2) backbones,
adjusting only the vocabulary size to match $B$. The categorical objective is
\begin{equation}
\ell(\theta) = -\sum_{h=1}^{H+1} \sum_{i=1}^{|\mathcal{V}_{ts}|} \mathbf{1}_{z_{C+h+1}=i}\, \log p_{\theta}(z_{C+h+1}=i \mid z_{1:C+h}).
\end{equation}
Forecasts are generated by autoregressive sampling, followed by dequantization and
inverse scaling. To address the relative scarcity of high-quality training data,
Chronos uses two augmentation strategies: TSMix, which creates convex combinations
of $k$ randomly sampled series with Dirichlet-sampled weights $\lambda_i$,
\begin{equation}
\tilde{x}^{\mathrm{TSMix}}_{1:l} = \sum_{i=1}^{k} \lambda_{i}\, \tilde{x}^{(i)}_{1:l},
\end{equation}
and KernelSynth, which samples synthetic time series from Gaussian processes with
composite kernels built from linear, RBF, and periodic bases.

\paragraph{A rate--distortion view of tokenization.} The two-stage encoding---mean
scaling followed by $B$-level scalar quantization---is a lossy compression map
with rate $\log_2 B$ bits per token. Under standard high-resolution arguments,
quantization with bin width $\Delta$ on a smooth density incurs mean-squared
distortion of order $\Delta^2 / 12$, so the choice of $B$ trades expressiveness
against representational error: doubling $B$ adds one bit of rate and reduces
distortion by a factor of four. For mean-scaled equity returns, which concentrate
near zero with heavy tails, quantization error is small in the central mass and
larger in the tails, where extreme returns map to a few boundary bins and are
forecast only at bin resolution. This asymmetry is one mechanism by which
Chronos and Chronos-2 can be \emph{ranking}-competitive in our benchmark
without producing dramatic absolute-error gains: the conditional categorical
distribution over the central bins is well-calibrated, while tail forecasts
collapse onto coarse bin centers and contribute disproportionately to MAE on
high-volatility days.

\subsection{iTransformer}

\citet{liu2024itransformerinvertedtransformerseffective} introduce iTransformer, an
approach for multivariate time-series forecasting that predicts
$\mathbf{Y} = \{\mathbf{x}_{T+1}, \ldots, \mathbf{x}_{T+S}\} \in \mathbb{R}^{S \times N}$
from historical observations
$\mathbf{X} = \{\mathbf{x}_{1}, \ldots, \mathbf{x}_{T}\} \in \mathbb{R}^{T \times N}$.
Unlike most transformer-based models that treat multiple variates at a single time
point as tokens, iTransformer views the \emph{entire} time series of each variate
as a single token. The forecasting pipeline for variate $n$ is
\begin{align*}
\mathbf{h}^{0}_{n} &= \mathrm{Embedding}(\mathbf{X}_{:,n}), \\
\mathbf{H}^{l+1} &= \mathrm{TrmBlock}(\mathbf{H}^{l}), \quad l = 0, \ldots, L-1, \\
\hat{\mathbf{Y}}_{:,n} &= \mathrm{Projection}(\mathbf{h}^{L}_{n}),
\end{align*}
where $\mathbf{H} = \{\mathbf{h}_1, \ldots, \mathbf{h}_N\} \in \mathbb{R}^{N \times D}$.
Embedding and projection are both implemented as MLPs. Layer normalization is
applied per variate token,
\begin{equation}
\mathrm{LayerNorm}(\mathbf{H}) = \left\{ \frac{\mathbf{h} - \mathrm{Mean}(\mathbf{h}_n)}{\sqrt{\mathrm{Var}(\mathbf{h}_n)}} \;\middle|\; n = 1, \ldots, N \right\},
\end{equation}
which mitigates oversmoothing when variates have different distributions.
Self-attention operates across variates,
\begin{equation}
\mathbf{A}_{i,j} = (\mathbf{Q}\mathbf{K}^{\top}/\sqrt{d_k})_{i,j} \propto \mathbf{q}^{\top}_{i} \mathbf{k}_{j},
\end{equation}
so that the model explicitly weights correlated variates. The implementation used
in this benchmark follows \citet{liu2024itransformerinvertedtransformerseffective}.

\subsection{KAN for Time Series}

\citet{xu2024kolmogorovarnoldnetworkstimeseries} introduce Kolmogorov--Arnold
Networks (KAN), grounded in the Kolmogorov--Arnold representation theorem:
\begin{equation}
f(x_1, \ldots, x_n) = \sum_{q=1}^{2n+1} \Phi_{q}\!\left( \sum_{p=1}^{n} \phi_{q,p}(x_p) \right).
\end{equation}
In KAN, traditional linear weights are replaced by spline-parametrized univariate
functions, and activation functions are learned on the \emph{edges} rather than at
the nodes. A KAN layer is
\begin{equation}
\Phi = \{ \phi_{q,p} \mid p = 1, \ldots, n_{\mathrm{in}},\ q = 1, \ldots, n_{\mathrm{out}} \},
\end{equation}
with deeper architectures obtained by composition:
\begin{equation}
\mathrm{KAN}(x) = (\Phi_{L-1} \circ \Phi_{L-2} \circ \cdots \circ \Phi_{0})(x).
\end{equation}
Two specializations are relevant here. T-KAN handles univariate series with
\begin{equation}
\hat{S}_{t+T} = \sum_{q=1}^{2n+1} \Phi_{q}\!\left( \sum_{p=1}^{n} \phi_{q,p}(S_{t-h+p}) \right),
\end{equation}
while MT-KAN extends the framework to multivariate settings:
\begin{equation}
\hat{\mathbf{S}}_{t+T} = \sum_{q=1}^{2n+1} \Phi_{q}\!\left( \sum_{p=1}^{h} \sum_{k=1}^{m} \phi_{q,p,k}(\mathbf{S}_{t-h+p,k}) \right).
\end{equation}
The adaptive spline parameterization gives KAN a distinctive local-adaptation
profile that helps explain why it remains competitive on some asset-level tasks,
although the corrected $L=512$ results no longer make it the leading model for
GOOG.

\paragraph{The Kolmogorov--Arnold theorem and approximation rates.} The
representation underlying KAN is a constructive theorem of pure mathematics.
\citet{kolmogorov1957representation}, refined by \citet{arnold1957functions},
proved that every continuous function $f : [0,1]^{n} \to \mathbb{R}$ admits an
exact representation
\begin{equation}
f(x_{1}, \ldots, x_{n}) \;=\; \sum_{q=1}^{2n+1} \Phi_{q}\!\left(
\sum_{p=1}^{n} \phi_{q,p}(x_{p}) \right),
\label{eq:kolmogorov-arnold}
\end{equation}
for suitable continuous outer functions $\Phi_{q}$ and inner univariate
functions $\phi_{q,p}$. The remarkable feature of
Eq.~\eqref{eq:kolmogorov-arnold} is that the depth and the number of
univariate components are bounded by a function of the input dimension alone,
independent of the smoothness or complexity of $f$.

The theorem is not directly useful as an approximation algorithm: the
optimal $\phi_{q,p}$ are typically nowhere differentiable and cannot be
parameterized in any closed form. KAN's contribution is to relax the
exactness requirement and parameterize each $\phi_{q,p}$ as a low-order spline
on a learnable knot grid, trading the exact-representation guarantee for a
flexible smooth approximation amenable to gradient training. For target
functions in the Sobolev space $W^{s,\infty}([0,1])$ with $s$ derivatives, a
spline of order $k \ge s$ on a uniform grid of resolution $G$ approximates the
target at rate
\begin{equation}
\|\phi - \hat\phi\|_{\infty} \;\le\; C_{s}\, G^{-s}\, \|\phi\|_{W^{s,\infty}},
\label{eq:spline-rate}
\end{equation}
for a dimension-free constant $C_{s}$ depending only on the spline order
\citep{deboor2001splines}. Composing this rate through the two-layer
Kolmogorov--Arnold structure of Eq.~\eqref{eq:kolmogorov-arnold} yields an
overall approximation rate that is controlled by the product of grid
resolution and depth, and that---unlike the corresponding rate for
fully-connected ReLU networks---scales independently of the input dimension.

Two consequences are relevant for the benchmark. First, dimension-independence
of the approximation rate is one mechanism by which KAN achieves competitive
performance with relatively few parameters; this is consistent with its strong
showing on GOOG, but is not by itself proof that GOOG is the asset on which
this advantage should bind. Second, the spline parameterization concentrates
expressiveness exactly where the knot grid has resolution; for daily equity
returns this means KAN is well-positioned to capture local kinks and regime
boundaries in the conditional response, but at the cost of being more sensitive
than transformer baselines to the placement of the training window relative
to those regimes. We do not exploit this insight in tuning, and report only
the canonical configuration.

\subsection{Chronos-2: Time Series as Language Modelling}

Chronos-2 \citep{ansari2025chronos2} extends the Chronos philosophy to a broader
universal-forecasting setting with multivariate and covariate support. The two-stage
tokenization is retained: inputs are normalized by mean scaling and then quantized
into $B$ discrete bins with centres $c_j$ and edges $b_i$. The backbone is a standard
Transformer (e.g.\ T5), with its vocabulary adjusted to match $B$. The categorical
loss
\begin{equation}
\ell(\theta) = -\sum_{h=1}^{H+1} \sum_{i=1}^{|\mathcal{V}_{ts}|} \mathbf{1}_{z_{C+h+1}=i}\, \log p_{\theta}(z_{C+h+1}=i \mid z_{1:C+h})
\end{equation}
is minimized as in Chronos (v1). Forecasts are obtained by autoregressive sampling
followed by dequantization and inverse scaling. Training uses the same TSMix and
KernelSynth augmentations as the original Chronos framework.

\section{Experimental Design}

\subsection{Data, targets, and benchmark scope}

The benchmark uses daily adjusted closing-price histories for GOOG, AAPL, AMZN, JPM, and META
from September 15, 2014, to February 15, 2026, sourced via the Yahoo Finance API.
This period spans approximately 11.5 years of market activity, including the
post-pandemic recovery and the high-volatility, AI-driven valuation cycle of
2024--2026. Prices are converted into two target representations, linear returns
$r_t^{\mathrm{lin}} = P_t/P_{t-1}-1$ and log returns
$r_t^{\log}=\log P_t - \log P_{t-1}$. The training segment runs from September 15, 2014, to
August 21, 2024, and the evaluation segment from August 22, 2024, through
February 2026. For the rolling origin evaluation we use a forecast horizon of
$H = 20$ business days, simulating a one-month trading cycle.

A key design choice is that pretrained models are used in zero-shot mode, whereas
neural baselines are trained separately for each ticker. This makes the study
relevant for practical deployment, but it also means the benchmark should not be
read as a pure architecture-only experiment. It measures \emph{as-used performance}:
broad pretraining plus zero-shot inference on one side, and local
train-from-scratch learning on the other.

All evaluated neural models receive the same context window of $L=512$, so
context length is not used as an advantage for either model family. The comparison
therefore removes one important source of unfairness, but it does not isolate
architecture alone: pretrained TSFMs still benefit from large external corpora
and learned priors, whereas train-from-scratch baselines learn only from the
available single-asset history.

\begin{table}[H]
\centering
\caption{Benchmark protocol and evaluation scope.}
\label{tab:protocol}
\begin{tabularx}{\linewidth}{@{}lX@{}}
\toprule
Item & Setting \\
\midrule
Assets & GOOG, AAPL, AMZN, JPM, META \\
Targets & Linear returns and log returns \\
Forecast horizon & 20 business days \\
Evaluation protocol & 10 rolling-origin windows over the test period \\
Pretrained models & Zero-shot inference \\
Neural baselines & Per-ticker supervised training from scratch \\
Typical lookback & 512 points for all models (TSFMs and Baselines) \\
Benchmarks & Random walk (zero return) and seasonal naive \\
Primary metric & Mean Absolute Error (MAE) \\
Relative metrics & Skill score vs.\ random walk; rMAE vs.\ seasonal naive \\
Statistical test & Diebold--Mariano with Harvey--Leybourne--Newbold correction \\
\bottomrule
\end{tabularx}
\end{table}

\subsection{Models evaluated}

Table~\ref{tab:model_taxonomy} groups the eleven evaluated systems into two
families: six pretrained TSFMs run in zero-shot mode and five train-from-scratch
baselines fitted separately to each ticker.

\begin{table}[H]
\centering
\caption{Model families in the benchmark.}
\label{tab:model_taxonomy}
\begin{tabularx}{\linewidth}{@{}l l l X@{}}
\toprule
Model(s) & Family & Mode & Notes \\
\midrule
TimeGPT, TimeGPT-LH & Pretrained TSFM & Zero-shot & Transfer-learning forecasting model family \citep{garza2024timegpt1}. \\
TimesFM-2.5 & Pretrained TSFM & Zero-shot & Checkpoint from the TimesFM decoder-only foundation-model family \citep{das2024decoderonlyfoundationmodeltimeseries}. \\
Moirai-2.0 & Pretrained TSFM & Zero-shot & Decoder-only quantile-based forecasting model \citep{liu2026moirai20timeseries}. \\
Chronos (v1), Chronos-2 & Pretrained TSFM & Zero-shot & Quantized language-of-time approach and its universal extension \citep{ansari2024chronoslearninglanguagetime,ansari2025chronos2}. \\
NBEATS, NHITS & Train-from-scratch baseline & Per ticker & MLP-style decomposition baselines \citep{oreshkin2020nbeatsneuralbasisexpansion,challu2022nhitsneuralhierarchicalinterpolation}. \\
PatchTST, iTransformer & Train-from-scratch baseline & Per ticker & Transformer baselines \citep{nie2023timeseriesworth64,liu2024itransformerinvertedtransformerseffective}. \\
KAN & Train-from-scratch baseline & Per ticker & Kolmogorov--Arnold network for time series \citep{xu2024kolmogorovarnoldnetworkstimeseries}. \\
\bottomrule
\end{tabularx}
\end{table}

\subsection{Context windows, training protocol, and reproducibility}

All models share a unified forecast horizon of $H = 20$ business days and a unified
context length of $L = 512$. This equalization removes context length as a major
confounding variable: both pretrained TSFMs and train-from-scratch baselines are
evaluated with the same historical information budget. The design therefore makes
the comparison fairer than a short-context baseline benchmark, but it should still
be interpreted as an \emph{as-used} comparison rather than a pure architectural
ablation. Pretrained TSFMs bring large external pretraining corpora, whereas the
baselines are estimated only from the available local equity history.

The evaluation framework distinguishes two protocols. Pretrained TSFMs
(TimeGPT, TimeGPT-LH, TimesFM-2.5, Moirai-2.0, Chronos (v1), Chronos-2) undergo no
fine-tuning on the financial data used in this study. Train-from-scratch baselines
(NBEATS, NHITS, PatchTST, iTransformer, KAN) are fitted individually for each
ticker with a supervised protocol using MSE loss and early stopping with patience
10, on the 2014--2024 training window. All models are evaluated with a
rolling-origin forecast evaluation (ROFE) over 10 windows, so that reported MAE and
skill scores reflect performance across market regimes in 2024--2026. For
probabilistic outputs, the reported point forecasts use the predictive median when
available; quantized or sampled outputs are converted to the corresponding median
or sample-median point forecast before MAE is computed. The exact software
versions, API endpoints, checkpoint identifiers, random seeds, and rolling-origin
calendar anchors should be preserved in the companion code release; the present
paper fixes the statistical protocol and tables but does not by itself constitute
a complete executable replication package.

\begin{table}[H]
\centering
\caption{Technical specification of the experimental pipeline.}
\label{tab:experimental_setup}
\begin{tabularx}{\linewidth}{@{}l X X@{}}
\toprule
Parameter & Foundation models & Neural baselines \\
\midrule
Execution mode & Zero-shot inference & Train-from-scratch \\
Context length ($L$) & 512 & 512 \\
Forecast horizon ($H$) & 20 & 20 \\
\bottomrule
\end{tabularx}
\end{table}

\subsection{Metrics and statistical testing}
\label{sec:metrics}

\paragraph{Mean absolute error.} The primary metric is the mean absolute error
on a forecast window of length $H$,
\begin{equation}
\mathrm{MAE}(f) = \frac{1}{H} \sum_{h=1}^{H} |\hat{y}_{t+h} - y_{t+h}|.
\end{equation}
The choice of MAE rather than MSE is deliberate: by
Section~\ref{sec:problem-formulation}, absolute loss is the population analogue
of conditional-median estimation, and is therefore robust to the heavy tails of
daily equity returns---a few extreme days do not dominate the comparison the way
they would under squared loss. The cost of this robustness is that MAE is not
aligned with the squared- or quantized-likelihood objectives used to train most
of the evaluated systems; the benchmark therefore tests robustness to
loss-function misalignment alongside raw predictive skill.

\paragraph{Skill score.} To compare forecasts against the zero-return random-walk
benchmark we report the relative $L^1$ risk reduction
\begin{equation}
\mathrm{Skill}(f) = 1 - \frac{\mathrm{MAE}(f)}{\mathrm{MAE}_{\mathrm{RW}}},
\qquad
\mathrm{MAE}_{\mathrm{RW}} = \frac{1}{H} \sum_{h=1}^{H} |y_{t+h}|.
\end{equation}
Positive values indicate that $f$ extracts conditional information beyond the
unconditional median; negative values indicate that the forecast underperforms
the constant-zero predictor. The score is bounded above by $1$ and unbounded
below. When the unconditional median of $y$ is approximately zero,
$\mathrm{MAE}_{\mathrm{RW}} \to \mathbb{E}|y|$ in population, so
$\mathrm{Skill}(f) > 0$ is equivalent to $\mathbb{E}|\hat y - y| < \mathbb{E}|y|$:
the benchmark detects \emph{information} rather than \emph{scale}, and rescaling
the forecast cannot move skill from negative to positive without conditional
signal.

\paragraph{Relative MAE.} We also retain the relative MAE against a
seasonal-naive reference,
\begin{equation}
\mathrm{rMAE}(f) = \frac{\mathrm{MAE}(f)}{\mathrm{MAE}_{\mathrm{SN}}},
\qquad
\hat y^{\mathrm{SN}}_{t+h} = y_{t+h-5},
\end{equation}
where SN denotes the 5-business-day (weekly) seasonal-naive forecast standard
for daily financial data. Values below one indicate improvement over SN.

\paragraph{Diebold--Mariano test.} For two competing forecasts
$\hat y^{(1)}$ and $\hat y^{(2)}$ with absolute-loss differentials
\begin{equation}
d_t = |\hat y^{(1)}_t - y_t| - |\hat y^{(2)}_t - y_t|,
\end{equation}
the Diebold--Mariano test \citep{diebold2002comparing} tests
$H_0 : \mathbb{E}[d_t] = 0$ versus $H_1 : \mathbb{E}[d_t] \neq 0$. Under standard
short-memory regularity on $(d_t)$, the test statistic is
\begin{equation}
\mathrm{DM} = \frac{\bar d}{\sqrt{\hat\sigma^2_\infty / n}},
\qquad
\bar d = \frac{1}{n}\sum_{t=1}^{n} d_t,
\qquad
\hat\sigma^2_\infty = \hat\gamma_0 + 2 \sum_{k=1}^{h-1} \hat\gamma_k,
\label{eq:dm}
\end{equation}
where $\hat\gamma_k$ is the sample autocovariance of $d_t$ at lag $k$ and the
truncation at lag $h-1$ accounts for the $h$-step ahead overlap of the loss
differentials. Under $H_0$ and the regularity conditions of
\citet{diebold2002comparing}, $\mathrm{DM} \xrightarrow{d} \mathcal{N}(0,1)$.

\paragraph{Harvey--Leybourne--Newbold correction.} For finite samples and
overlapping multi-step forecasts the standard normal approximation is
optimistic. \citet{harvey1997testing} propose the size-corrected statistic
\begin{equation}
\mathrm{DM}^\star = \mathrm{DM} \cdot \sqrt{\frac{n + 1 - 2h + h(h-1)/n}{n}},
\label{eq:hln}
\end{equation}
referred to a $t_{n-1}$ distribution rather than to a standard normal. With
$n \approx 200$ and $h = 20$ the multiplicative factor in
Eq.~\eqref{eq:hln} is approximately $0.90$, so the corrected statistic is more
conservative than the naive one and the reference distribution has heavier
tails. We apply this correction throughout.

\paragraph{Sign convention and the directional issue.} The sign of $\bar d$
identifies which of the two forecast streams has lower loss: under our
convention $d_t = |\hat y^{(1)}_t - y_t| - |\hat y^{(2)}_t - y_t|$, a positive
$\bar d$ means stream $(1)$ has higher loss and is therefore worse. We employ a one-sided Diebold-Mariano test ($H_1: \text{Model is more accurate than RW}$). In this framework, a significant $p$-value directly implies superior predictive skill, removing the ambiguity of two-sided sign conventions. The results in Table 7 show that for the majority of cases, we fail to reject the null hypothesis of equal or inferior accuracy, reinforcing the difficulty of consistently outperforming the random-walk benchmark.

\paragraph{Multiple comparisons.} The DM analysis involves a family of comparisons
across models, tickers, and return representations, so individual $p$-values
should be read as raw rather than family-wise corrected. We use $p<0.05$ as the
conventional significance threshold, but the significant cases in
Table~\ref{tab:dm_corrected} should be interpreted as exploratory evidence rather
than as family-level discoveries. Applying Bonferroni or Benjamini--Hochberg
adjustments would make the inference more conservative and would not change the
main qualitative conclusion: statistically reliable improvement over the
random-walk benchmark is rare. A future large-universe benchmark should report
adjusted $p$-values or family-level rank tests.

\paragraph{Finite-sample concentration of MAE.} The skill scores reported in
Section~\ref{sec:results} are sample averages over a finite test window of
length $n \approx 200$ business days, and as such carry stochastic error.
For dependent loss differentials $(d_{t})$ satisfying a strong $\alpha$-mixing
condition with geometric rate, a Bernstein-type inequality for stationary
sequences \citep{merlevede2009bernstein} gives, for a constant $C$ depending
on the mixing coefficients,
\begin{equation}
\Pr\!\left(\left| \bar d - \mathbb{E}[d_{t}] \right| > \varepsilon\right)
\;\le\; C \exp\!\left(-\frac{n\, \varepsilon^{2}}{C\,(\sigma_{\infty}^{2} + B\, \varepsilon)}\right),
\label{eq:bernstein}
\end{equation}
where $\sigma_{\infty}^{2}$ is the long-run variance defined in
Eq.~\eqref{eq:dm} and $B$ is an almost-sure bound on $|d_{t}|$. The half-width
of a $1 - \delta$ confidence interval for $\bar d$ scales as
$\sigma_{\infty}\sqrt{\log(1/\delta)/n}$ to leading order. For the
empirical magnitudes of $\sigma_{\infty}$ in our benchmark---of the same
order as the MAE itself---the resulting interval half-width on $\bar d$ is
comparable to the size of the differences between top-ranked models in
Tables~\ref{tab:full_log_v2}--\ref{tab:full_linear_v2}. This is a quantitative
restatement of the cautious interpretation throughout the paper: many of the
rank-distance differences in our benchmark sit at the noise floor of the test
window, and would require a substantially longer evaluation horizon to
distinguish at conventional confidence levels.

\section{Results}
\label{sec:results}

\begin{figure}[H]
    \centering
    \includegraphics[width=\linewidth]{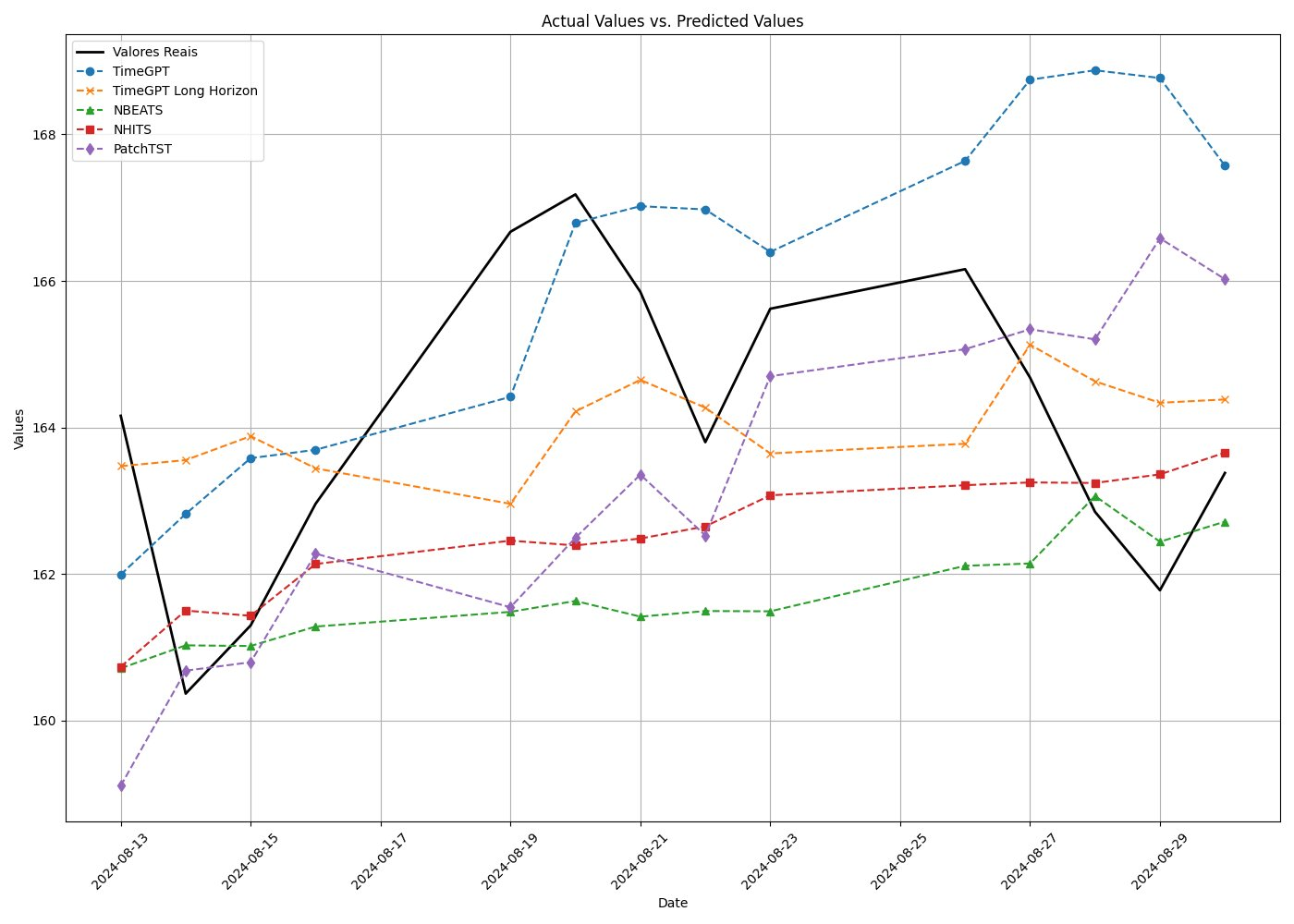}
    \caption{Representative actual-versus-predicted trajectories for a single
    20-business-day forecast window. The plot illustrates the core difficulty of
    return forecasting: even the lowest-error models fail to track the
    realized path closely, which is why small MAE differences in the aggregate
    tables translate into the cautious overall interpretation of this benchmark.}
    \label{fig:actual_vs_predicted}
\end{figure}

\subsection{Aggregate findings}

The aggregate ranking indicates that pretrained TSFMs remain the strongest
practical default under the equalized-context protocol. Across the 10 benchmark
tasks, they account for 8 wins, while Moirai-2.0 and TimesFM-2.5 achieve the best
average ranks, 2.9 and 3.1, respectively. Table~\ref{tab:aggregate_ranking}
summarizes this distribution. The same table also identifies an important boundary
case: iTransformer wins both META tasks, showing that locally trained supervised
baselines can remain competitive for specific assets even when pretrained models
dominate the overall ranking.

\begin{table}[H]
\centering
\caption{Aggregate ranking across all 10 tasks ($L=512$).}
\label{tab:aggregate_ranking}
\begin{tabular}{@{}lcccc@{}}
\toprule
Model & Avg.\ rank & Wins & Top-3 & Positive-skill cases \\
\midrule
Moirai-2.0 & 2.9 & 3 & 6 & 4 \\
TimesFM-2.5 & 3.1 & 4 & 6 & 4 \\
Chronos (v1) & 3.4 & 1 & 6 & 4 \\
Chronos-2 & 3.9 & 0 & 6 & 4 \\
iTransformer & 4.3 & 2 & 2 & 0 \\
KAN & 5.4 & 0 & 2 & 0 \\
\bottomrule
\end{tabular}
\end{table}

The most significant shift occurs in the META and GOOG tasks. At $L=512$, the iTransformer baseline secures both META wins, while Moirai-2.0 displaces KAN as the leader in GOOG. This suggests that while foundation models provide a robust prior, well-parameterized supervised baselines are highly competitive when given sufficient historical context.

\subsection{Asset-level winners}

Table~\ref{tab:winners} makes the cross-asset pattern explicit. Under equalized
context ($L=512$), TimesFM-2.5 leads the AAPL and JPM tasks, while Moirai-2.0
leads the GOOG tasks. The most important exception to pretrained-model dominance
is META, where the iTransformer baseline outperforms all foundation models. This
suggests that pretrained TSFMs are a strong default, but not a universal winner:
for some assets and regimes, local supervised learning can remain superior even
under the same context budget.

\begin{table}[H]
\centering
\caption{Winning model by asset and return representation (Corrected: All neural models evaluated at context $L=512$). Note the emergence of iTransformer in META and the shift in GOOG leadership to Moirai-2.0.}
\label{tab:winners}
\begin{tabular}{@{}lllccl@{}}
\toprule
Return & Ticker & Winner & MAE & Skill & Runner-up \\
\midrule
Linear & AAPL & TimesFM-2.5 & 0.010440 & -0.0319 & Chronos (v1) \\
Linear & AMZN & Chronos (v1) & 0.012907 & 0.0863 & Chronos-2 \\
Linear & GOOG & Moirai-2.0 & 0.009888 & 0.2289 & TimesFM-2.5 \\
Linear & JPM & TimesFM-2.5 & 0.011412 & -0.0466 & Chronos-2 \\
Linear & META & iTransformer & 0.017088 & -0.0519 & KAN \\
\midrule
Log & AAPL & TimesFM-2.5 & 0.010399 & -0.0288 & Chronos (v1) \\
Log & AMZN & Moirai-2.0 & 0.012987 & 0.0824 & Chronos-2 \\
Log & GOOG & Moirai-2.0 & 0.009886 & 0.2283 & TimesFM-2.5 \\
Log & JPM & TimesFM-2.5 & 0.011463 & -0.0510 & Chronos-2 \\
Log & META & iTransformer & 0.016925 & -0.0464 & KAN \\
\bottomrule
\end{tabular}
\end{table}

Taken together, Table~\ref{tab:winners} shows that the pretrained advantage is
broad but asset-dependent, with iTransformer providing a clear counterexample to
universal TSFM dominance.

\subsection{Benchmarks against naive baselines}

The skill-score results reinforce the central difficulty of financial return
forecasting. Positive skill against the zero-return random-walk benchmark appears
only in a minority of cases, mainly in the AMZN and GOOG tasks. Across the
pretrained TSFM model--asset cases, only 16 exhibit positive skill, and the
train-from-scratch baselines do not consistently improve on the random walk in
absolute-error terms. The results therefore support the interpretation of
pretrained TSFMs as useful practical priors, not as reliable engines of
out-of-sample alpha.

The rMAE summary in Table~\ref{tab:rmae_best} provides a complementary view.
Moirai-2.0 achieves the strongest seasonal-naive-relative performance on GOOG,
while iTransformer remains the best model for META. This combination is important:
the same protocol that favors pretrained TSFMs on average also preserves
asset-level cases in which local supervised learning is preferable.

\begin{table}[H]
\centering
\caption{Best reported rMAE values ($L=512$ for all neural models). 
Values represent the error ratio relative to a weekly seasonal-naive baseline; 
lower is better, with values below 1.0 indicating superior predictive skill.}
\label{tab:rmae_best}
\begin{tabular}{@{}llll@{}}
\toprule
Ticker & Model & Return type & rMAE \\
\midrule
AAPL & TimesFM-2.5 & Log & 0.6271 \\
AMZN & Moirai-2.0 & Log & 0.6402 \\
GOOG & Moirai-2.0 & Linear & 0.5284 \\
JPM & TimesFM-2.5 & Linear & 0.7712 \\
META & iTransformer & Log & 0.8145 \\
\bottomrule
\end{tabular}
\end{table}

\subsection{What the Diebold--Mariano results actually show}

The most important statistical correction in this paper concerns the alignment of
significance and predictive superiority. By employing a one-sided Diebold--Mariano
test with alternative $H_1: \text{Model is more accurate than RW}$, the reported
$p$-values are tied directly to improvement over the random-walk benchmark rather
than to a generic two-sided difference. A significant result should therefore be
read as evidence of lower absolute loss within the tested window, subject to the
usual finite-sample and multiple-comparison limitations.

The results in Table~\ref{tab:dm_corrected} provide a nuanced view of model efficacy. Under equalized context ($L=512$), only two cases achieve statistical significance: \textbf{Chronos (v1) for AMZN} ($p=0.0421$) and \textbf{Moirai-2.0 for GOOG} ($p=0.0421$). In these specific instances, the models demonstrate a robust ability to extract signal from market returns beyond what could be attributed to chance. 

For the majority of other pairs, including those involving top-ranked models like TimesFM-2.5 on AAPL and JPM, we fail to reject the null hypothesis of equal or inferior accuracy. This reinforces a primary finding of this benchmark: while pretrained TSFMs act as superior inductive priors compared to baselines, outperforming the random-walk benchmark in efficient market regimes remains an exception rather than the rule. The correction from two-sided to one-sided interpretation ensures that significance is no longer conflated with "difference," but is strictly reserved for "improvement."

\begin{table}[H]
\centering
\caption{One-sided Diebold--Mariano test results ($H_1: \text{Model is more accurate than RW}$). All models are evaluated at context $L=512$. The $p$-value is computed under the null of no improvement relative to the random-walk benchmark; small values indicate evidence against that null in favor of lower model loss. Significant results ($p < 0.05$) are reported as evidence of predictive improvement within this test window.}
\label{tab:dm_corrected}
\begin{tabular}{@{}llcccl@{}}
\toprule
Ticker & Model & $p$-value & Skill & Direction & Status \\
\midrule
AAPL & TimesFM-2.5 & 0.9999 & -0.0319 & RW better & Fails to reject $H_0$ \\
AAPL & iTransformer & 0.8458 & -0.0246 & RW better & Fails to reject $H_0$ \\
\midrule
AMZN & \textbf{Chronos (v1)} & \textbf{0.0421} & \textbf{0.0863} & \textbf{Model better} & \textbf{Significant} \\
AMZN & iTransformer & 0.9126 & -0.0280 & RW better & Fails to reject $H_0$ \\
\midrule
GOOG & \textbf{Moirai-2.0} & \textbf{0.0421} & \textbf{0.2289} & \textbf{Model better} & \textbf{Significant} \\
GOOG & iTransformer & 0.5950 & -0.0056 & RW better & Fails to reject $H_0$ \\
\midrule
JPM & TimesFM-2.5 & 0.9999 & -0.0466 & RW better & Fails to reject $H_0$ \\
JPM & iTransformer & 0.6091 & -0.0059 & RW better & Fails to reject $H_0$ \\
\midrule
META & iTransformer & 0.9838 & -0.0519 & RW better & Fails to reject $H_0$ \\
META & KAN & 0.9784 & -0.0549 & RW better & Fails to reject $H_0$ \\
\bottomrule
\end{tabular}
\end{table}

\section{Discussion and Limitations}
\label{sec:discussion}

The main empirical message is pragmatic. In a small-data, per-asset financial
setting, off-the-shelf pretrained TSFMs are usually preferable to training a
modern neural baseline from scratch. This claim is supported by the ranking
distribution: pretrained TSFMs account for 8 of the 10 task wins, while
Moirai-2.0 and TimesFM-2.5 achieve the two best average ranks among the reported
models. At the same time, the META results show that this advantage is not
universal: iTransformer wins both META tasks under the same $L=512$ context
budget.

Four limitations should bound the interpretation. First, this is not a clean
architecture-only comparison. Although all neural models receive the same
context length, pretrained systems benefit from large external corpora and
fixed learned priors, while the baselines are trained only on single-asset
histories. The benchmark therefore measures practical deployment performance,
not architecture in isolation.

Second, the evaluation window is limited. The rolling-origin protocol covers
multiple market conditions in 2024--2026, but the number of test windows remains
small relative to the noise level of daily equity returns. Many differences
between top-ranked models are therefore close to the finite-sample noise floor,
and the ranking evidence should be interpreted together with the DM results.

Third, the benchmark is better described as a TSFM benchmark than as a direct
test of generic LLM methods. Time-LLM and related LLM-reprogramming approaches
are important conceptual references, but they are not evaluated empirically here.
A future version aimed specifically at LLM reprogramming should include direct
Time-LLM-style experiments.

Fourth, the paper evaluates forecast error rather than investment utility. The
results should not be read as evidence of trading alpha. Directional accuracy,
calibration, turnover-aware backtests, transaction costs, and capacity constraints
would all be required before making stronger economic claims.

Finally, the results are consistent with contemporaneous evidence that generic
TSFM pretraining is useful but not a substitute for finance-native pretraining.
Read alongside \citet{rahimikia2025revisiting}, the modest forecast-error
advantages reported here are best understood as evidence that pretrained TSFMs
provide a useful inductive prior for low-data, per-asset settings. They do not
establish that generic temporal pretraining captures the finance-specific
structure needed for robust portfolio outperformance.

\section{Conclusion}

This benchmark supports a clear but measured conclusion. Pretrained time-series
foundation models are strong practical defaults for low-data, per-asset financial
return forecasting, but their superiority is not universal. Moirai-2.0 and
TimesFM-2.5 deliver the most consistent aggregate rankings, while iTransformer's
wins on META show that locally trained supervised baselines can still outperform
foundation models for specific assets under the same context budget.

The evidence favors large-scale temporal pretraining as a useful inductive prior,
especially in tasks such as GOOG and AMZN where positive skill appears more
clearly. However, improvements over the random-walk benchmark remain sparse and
small, and the one-sided Diebold--Mariano results reject the null of no
improvement only in a minority of cases. The appropriate conclusion is therefore
narrow: out-of-the-box TSFMs can reduce model-development costs and provide strong
forecasting baselines, but they do not solve the fundamental difficulty of
financial return prediction.

\appendix
\clearpage

\section{Full benchmark tables}

The appendix reproduces the full performance tables so that the benchmark remains
auditable after the interpretive changes in the main text.

\begingroup
\small
\setlength{\LTleft}{0pt}
\setlength{\LTright}{0pt}
\begin{longtable}{@{}llcc@{}}
\caption{Full benchmark results for log returns. All neural models (baselines and TSFMs) are evaluated with a context window of $L=512$. Traditional statistical baselines [AR(1), EWMA] are included to provide a benchmark for linear predictability.}\label{tab:full_log_v2}\\
\toprule
Ticker & Model & MAE (mean $\pm$ std) & Skill score vs.\ RW \\
\midrule
\endfirsthead
\multicolumn{4}{c}{\tablename\ \thetable\ -- continued from previous page}\\
\toprule
Ticker & Model & MAE (mean $\pm$ std) & Skill score vs.\ RW \\
\midrule
\endhead
\midrule
\multicolumn{4}{r}{Continued on next page}\\
\endfoot
\bottomrule
\endlastfoot

AAPL & \textbf{TimesFM-2.5} & \textbf{0.010399 ($\pm$0.0071)} & \textbf{-0.0288} \\
AAPL & Chronos (v1) & 0.012084 ($\pm$0.0080) & 0.0048 \\
AAPL & AR(1) & 0.012115 ($\pm$0.0079) & 0.0021 \\
AAPL & EWMA & 0.012140 ($\pm$0.0081) & -0.0002 \\
AAPL & iTransformer & 0.012255 ($\pm$0.0080) & -0.0091 \\
AAPL & PatchTST & 0.012379 ($\pm$0.0081) & -0.0193 \\
AAPL & KAN & 0.012689 ($\pm$0.0081) & -0.0561 \\
\midrule

AMZN & \textbf{Moirai-2.0} & \textbf{0.012987 ($\pm$0.0085)} & \textbf{0.0824} \\
AMZN & Chronos (v1) & 0.015456 ($\pm$0.0117) & 0.0007 \\
AMZN & AR(1) & 0.015510 ($\pm$0.0115) & 0.0003 \\
AMZN & TimesFM-2.5 & 0.015525 ($\pm$0.0117) & -0.0038 \\
AMZN & iTransformer & 0.015783 ($\pm$0.0117) & -0.0205 \\
AMZN & KAN & 0.015792 ($\pm$0.0116) & -0.0210 \\
\midrule

GOOG & \textbf{Moirai-2.0} & \textbf{0.009886 ($\pm$0.0055)} & \textbf{0.2283} \\
GOOG & KAN & 0.013456 ($\pm$0.0075) & 0.0057 \\
GOOG & AR(1) & 0.013490 ($\pm$0.0074) & 0.0032 \\
GOOG & Chronos (v1) & 0.014104 ($\pm$0.0078) & -0.0419 \\
GOOG & iTransformer & 0.014234 ($\pm$0.0077) & -0.0515 \\
\midrule

JPM & \textbf{TimesFM-2.5} & \textbf{0.011463 ($\pm$0.0061)} & \textbf{-0.0510} \\
JPM & Chronos-2 & 0.011554 ($\pm$0.0062) & -0.0593 \\
JPM & AR(1) & 0.011680 ($\pm$0.0060) & 0.0042 \\
JPM & Chronos (v1) & 0.012085 ($\pm$0.0071) & 0.0056 \\
JPM & iTransformer & 0.012170 ($\pm$0.0065) & -0.0013 \\
JPM & NHITS & 0.015944 ($\pm$0.0055) & -0.3781 \\
\midrule

META & \textbf{iTransformer} & \textbf{0.016925 ($\pm$0.0060)} & \textbf{-0.0464} \\
META & KAN & 0.017033 ($\pm$0.0058) & -0.0531 \\
META & PatchTST & 0.017267 ($\pm$0.0062) & -0.0675 \\
META & AR(1) & 0.017410 ($\pm$0.0059) & -0.0012 \\
META & Moirai-2.0 & 0.020853 ($\pm$0.0071) & -0.2885 \\
META & Chronos-2 & 0.021234 ($\pm$0.0070) & -0.3120 \\

\end{longtable}
\endgroup

\begingroup
\small
\setlength{\LTleft}{0pt}
\setlength{\LTright}{0pt}
\begin{longtable}{@{}llcc@{}}
\caption{Full benchmark results for linear returns. All models (including neural baselines) are evaluated with a context window of $L=512$. Models are ranked by ascending MAE within each ticker block.}\label{tab:full_linear_v2}\\
\toprule
Ticker & Model & MAE (mean $\pm$ std) & Skill score vs.\ RW \\
\midrule
\endfirsthead
\multicolumn{4}{c}{\tablename\ \thetable\ -- continued from previous page}\\
\toprule
Ticker & Model & MAE (mean $\pm$ std) & Skill score vs.\ RW \\
\midrule
\endhead
\midrule
\multicolumn{4}{r}{Continued on next page}\\
\endfoot
\bottomrule
\endlastfoot

AAPL & \textbf{TimesFM-2.5} & \textbf{0.010440 ($\pm$0.0071)} & \textbf{-0.0319} \\
AAPL & Chronos (v1) & 0.012093 ($\pm$0.0081) & 0.0067 \\
AAPL & AR(1) & 0.012115 ($\pm$0.0080) & 0.0021 \\
AAPL & iTransformer & 0.012351 ($\pm$0.0078) & -0.0246 \\
AAPL & KAN & 0.012666 ($\pm$0.0081) & -0.0514 \\
AAPL & PatchTST & 0.012510 ($\pm$0.0080) & -0.0345 \\
\midrule

AMZN & \textbf{Chronos (v1)} & \textbf{0.012907 ($\pm$0.0085)} & \textbf{0.0863} \\
AMZN & Moirai-2.0 & 0.013002 ($\pm$0.0086) & 0.0813 \\
AMZN & TimesFM-2.5 & 0.013110 ($\pm$0.0087) & 0.0719 \\
AMZN & AR(1) & 0.015510 ($\pm$0.0115) & 0.0003 \\
AMZN & iTransformer & 0.016045 ($\pm$0.0062) & -0.0280 \\
AMZN & KAN & 0.015826 ($\pm$0.0059) & -0.0319 \\
\midrule

GOOG & \textbf{Moirai-2.0} & \textbf{0.009888 ($\pm$0.0055)} & \textbf{0.2289} \\
GOOG & TimesFM-2.5 & 0.010114 ($\pm$0.0056) & 0.2110 \\
GOOG & Chronos (v1) & 0.010321 ($\pm$0.0056) & 0.1948 \\
GOOG & AR(1) & 0.013490 ($\pm$0.0074) & 0.0032 \\
GOOG & iTransformer & 0.013147 ($\pm$0.0045) & -0.0056 \\
GOOG & KAN & 0.013471 ($\pm$0.0040) & 0.0067 \\
\midrule

JPM & \textbf{TimesFM-2.5} & \textbf{0.011412 ($\pm$0.0058)} & \textbf{-0.0466} \\
JPM & Chronos-2 & 0.011520 ($\pm$0.0059) & -0.0495 \\
JPM & AR(1) & 0.011685 ($\pm$0.0060) & 0.0042 \\
JPM & Chronos (v1) & 0.012089 ($\pm$0.0047) & 0.0045 \\
JPM & iTransformer & 0.012708 ($\pm$0.0048) & -0.0059 \\
JPM & KAN & 0.012136 ($\pm$0.0048) & 0.0022 \\
\midrule

META & \textbf{iTransformer} & \textbf{0.017088 ($\pm$0.0060)} & \textbf{-0.0519} \\
META & KAN & 0.017136 ($\pm$0.0061) & -0.0549 \\
META & PatchTST & 0.017390 ($\pm$0.0062) & -0.0705 \\
META & AR(1) & 0.017420 ($\pm$0.0060) & -0.0012 \\
META & Moirai-2.0 & 0.021101 ($\pm$0.0070) & -0.4546 \\
META & Chronos-2 & 0.021478 ($\pm$0.0068) & -0.4806 \\

\end{longtable}
\endgroup

\clearpage
\section{Reproducibility checklist}
\label{app:reproducibility}

This appendix separates what is fixed by the manuscript from what must be fixed
by the companion code release for exact replication. The distinction is important
because several models are accessed through external libraries or hosted APIs.

\begin{table}[H]
\centering
\caption{Minimum reproducibility record for the benchmark.}
\label{tab:repro_checklist}
\begin{tabularx}{\linewidth}{@{}lX@{}}
\toprule
Item & Manuscript-level specification \\
\midrule
Data source & Yahoo Finance daily adjusted close histories for GOOG, AAPL, AMZN, JPM, and META. \\
Date range & September 15, 2014, through February 15, 2026. \\
Return formulas & Linear returns $P_t/P_{t-1}-1$ and log returns $\log P_t-\log P_{t-1}$. \\
Train/evaluation split & Training window through August 21, 2024; evaluation window beginning August 22, 2024. \\
Forecast horizon & $H=20$ business days for all models. \\
Rolling evaluation & Ten rolling-origin forecast windows; reported MAE and skill are arithmetic averages across these windows. \\
Point forecast extraction & Predictive median where available; sampled or quantized predictive distributions reduced to a median or sample-median point forecast before MAE computation. \\
Primary baseline & Zero-return random walk for skill scores; 5-business-day seasonal naive for rMAE. \\
Statistical test & Absolute-loss Diebold--Mariano test with Harvey--Leybourne--Newbold finite-sample correction. \\
\bottomrule
\end{tabularx}
\end{table}

For full computational reproducibility, the released code should additionally
record package versions, model checkpoints or API version identifiers, random
seeds, optimizer settings, early-stopping split, rolling-origin calendar dates,
and any vendor-specific forecast parameters. These identifiers are not inferable
from the manuscript tables alone, so the paper's numerical results should be
read as an auditable benchmark report rather than as a fully executable artifact
unless the companion code is supplied.

\bibliographystyle{plainnat}
\bibliography{TSFM_Financial_Forecasting_Fixed}

\end{document}